\documentclass[12pt]{article}

\usepackage{scicite}
\usepackage{graphicx}

\usepackage{times}

\newcommand{\solmas}{M$_{\odot}$}

\newenvironment{sciabstract}{%
\begin{quote} \bf}
{\end{quote}}

\newcounter{lastnote}

\title{The Formation and Fragmentation of Disks around Primordial Protostars}

\author
{Paul C. Clark,$^{1\ast}$ Simon C.O. Glover,$^{1}$ Rowan J. Smith,$^{1}$ Thomas H. Greif,$^{2}$ \\
Ralf S. Klessen,$^{1, 3}$ Volker Bromm$^{4}$\\
\\
\normalsize{$^{1}$ Institut f\"ur theoretische Astrophysik, Zentrum f\"ur Astronomie der Universit\"at Heidelberg,} \\ \normalsize{Albert-Ueberle-Str. 2, 69120, Heidelberg, Germany}\\
\normalsize{$^{2}$ Max-Planck-Institut f\"ur Astrophysik, Karl-Schwarzschild-Str. 1,} \\ 
\normalsize{D-85748 Garching, Germany}\\
\normalsize{$^{3}$ Kavli Institute for Particle Astrophysics and Cosmology,} \\ 
\normalsize{Stanford University, Menlo Park, CA 94025, USA} \\
\normalsize{$^{4}$ The University of Texas, Department of Astronomy and Texas Cosmology Center,} \\ 
\normalsize{2511 Speedway, RLM 15.306, Austin, TX 78712, USA} \\
\\
\normalsize{$^\ast$To whom correspondence should be addressed; E-mail:  pcc@ita.uni-heidelberg.de.}
}

\date{}

\begin{document} 


\baselineskip24pt


\maketitle


\begin{sciabstract}
%VB: small edits here!
 The very first stars to form in the universe heralded an end to the cosmic dark ages and introduced new physical processes that shaped early cosmic evolution. Until now, it was thought that these stars lived short, solitary lives, with only one extremely massive star, or possibly a very wide binary system, forming in each dark matter minihalo. Here we describe numerical simulations that show that these stars were, to the contrary, often members of tight multiple systems. Our results show that the disks that formed around the first young stars were unstable to gravitational fragmentation, possibly producing small binary and higher-order systems that had separations as small as the distance between the Earth and the Sun. 
\end{sciabstract}

\noindent The earliest stages of the formation of the first stars in the universe, often termed Population~III (Pop~III), have been well studied \cite{abn2002, bcl02, on2007}, with current numerical simulations evolving the collapsing gas from cosmological to protostellar densities \cite{yoha2006, yoh2008}. Much of the dynamical evolution of the gas during this phase is controlled by the formation of molecular hydrogen, the main coolant of the gas as it is dragged into the collapsing dark matter minihalos. The amount of H$_2$ formed sets the minimum gas temperature in the minihalos at around 200--300K, resulting in the first self-gravitating baryonic cores -- the initial conditions for primordial star formation -- having masses of around 1000 times that of the Sun (\solmas).

Until recently, it was assumed that each of these cores formed just a single star, because no fragmentation was seen in the simulations during the formation of the first protostar. As a result, attempts to estimate the final mass of the primordial stars have concentrated on balancing the inward accretion of gas from the collapsing core by the radiative feedback from the young protostar, with various calculations predicting a final mass in the range 30--300 \solmas \cite{abn2002, bl2004, yoh2008,mt2008}.

\begin{figure}
\centerline{
	\includegraphics[width=6.3in]{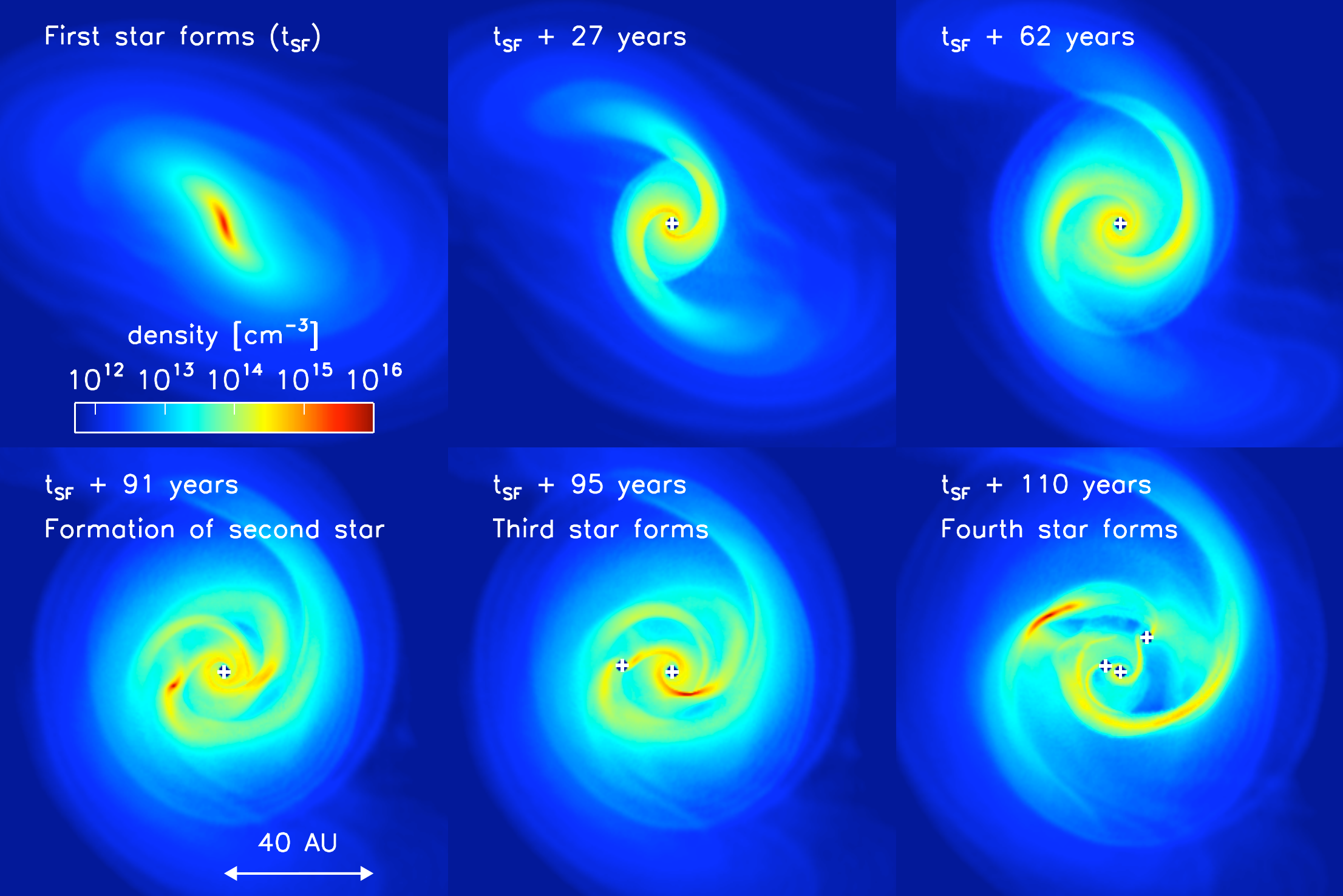}
}
\caption{Density evolution in a 120 AU region around the first protostar, showing the  build-up of the protostellar disk and its eventual fragmentation. We also see `wakes' in the low-density regions, produced by the previous passage of the spiral arms.}
\label{pcc_fig1}
\end{figure} 
 
It has been shown \cite{tao2009} that it is possible for the collapsing baryons to break into two distinct parts, each evolving independently to form its own star, thereby limiting the mass reservoir available for each component. The fragmentation of the gas arises from the chaotic turbulent flows that feed the inner regions of the star-forming minihalos. Numerical simulations suggest that this may occur in about one-fifth of all cases of primordial star formation \cite{tao2009}. However this figure is a lower limit, because the simulations were unable to follow the evolution of the gas beyond the formation of the initial protostar or binary system. A different study that followed the evolution of the gas at later times \cite{sgb2010} has shown that it settles into a disk with a radius of around 1000 astronomical units (AU), which is unstable to gravitational fragmentation. However, the limited mass resolution of this study, and the fact that it did not include the effects of the radiative feedback from the new-born stars, rendered its results inconclusive

Here, we present the results of a high resolution numerical simulation that captures the formation of the primordial protostar/disk system from cosmological initial conditions down to scales as small as 1.5 AU, and which includes the effects of the accretion luminosity heating as the disk builds up around the young protostar. We started by identifying the first dark matter minihalo to contain cooling, gravitationally collapsing gas from a simulation of a representative cosmological volume  \cite{SOM}.

We then re-zoomed the calculation using a technique called ``particle splitting'' [employed elsewhere in studies of primordial star formation, e.g. \cite{yoh2008}] and we focused our attention only on the collapsing gas at the center of the minihalo, ignoring the larger-scale evolution of the minihalo and its surroundings [see \cite{SOM} for details]. During this second stage of the simulation, the gas collapsed to very high densities. Normally, numerical simulations of this kind stop once the gas density exceeds around 10$^{14}$~cm$^{-3}$, because the computational cost of evolving the entire system beyond this point becomes prohibitively expensive. However, in our simulations, we replaced very high density collapsing regions with accreting `sink' particles \cite{SOM}, each of which represents an individual protostar. We then used the measured accretion rate onto the sink particles to calculate the luminosity produced by the mass as it fell onto the young protostar. This energy was then deposited into the surrounding gas under the assumption that the gas is optically thin, thus providing a conservative over-estimate of the heating from the protostar [see \cite{SOM} for details].

Prior to the formation of the first protostar, the results of our simulation were very similar to those presented elsewhere in the literature \cite{abn2002, yoh2008, tao2009}. However, the use of sink particles allows us to follow the evolution of the gas past the point at which the first protostar forms, and hence to simulate the build-up of an accretion disk around the protostar (Fig. 1). 

%We find that the disk is strongly self-gravitating, and highly susceptible to fragmentation, in spite of the heating due to the accretion luminosity from the central protostar. Since the formation of an accretion disk is a natural consequence of  the angular momentum present in the collapsing gas, we expect disk fragmentation and the formation of binaries and multiple systems to be a ubiquitous outcome of primordial star formation. Idealized calculations have previously found hints of such widespread binarity among the first stars \cite{machida08}, but our work establishes this property for realistic cosmological initial conditions, coupled with the radiative feedback, and the extremely high resolution needed to address protostellar disk evolution.

\begin{figure}
\centerline{
	\includegraphics[height=5.in]{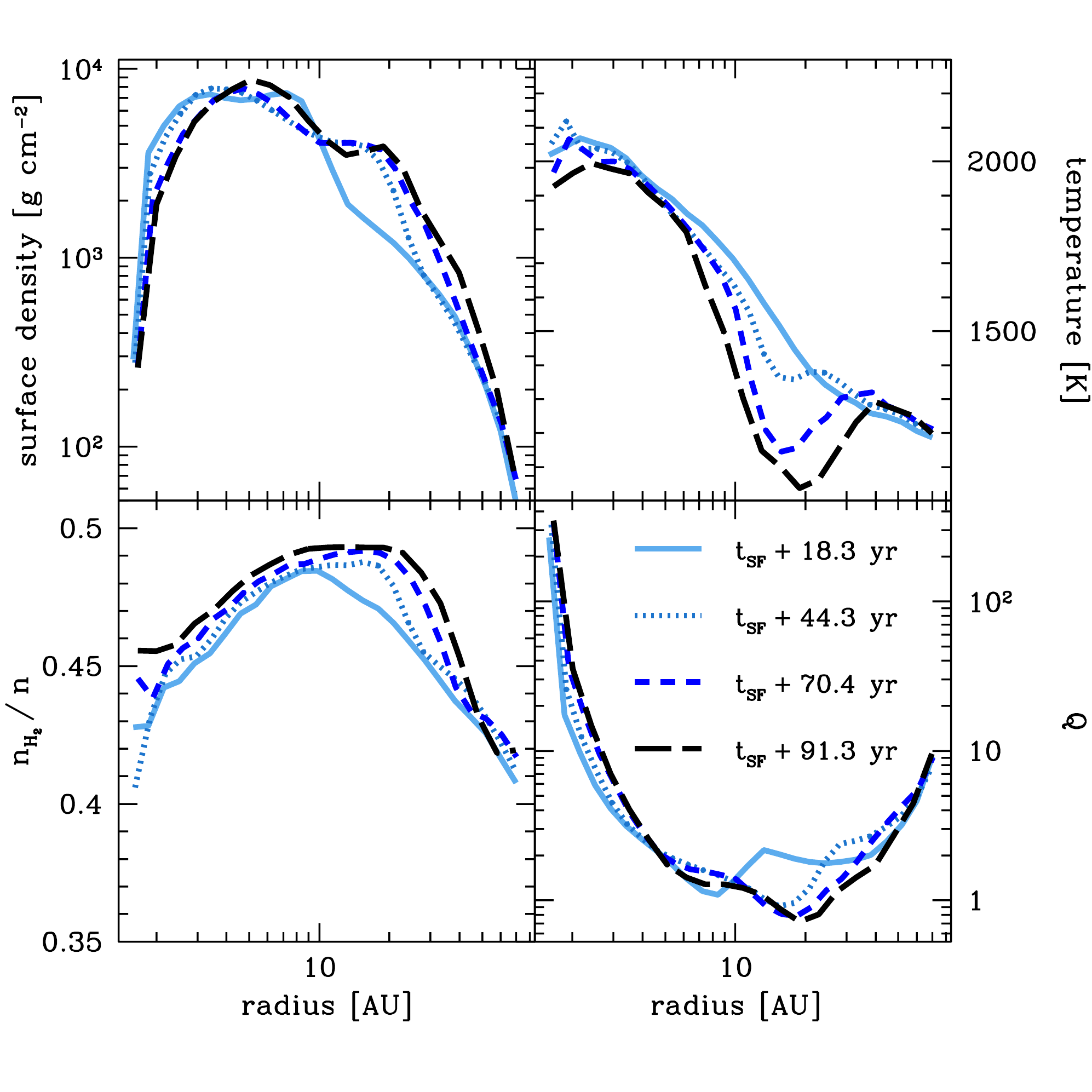}
}
\caption{Radial profiles of the disk's physical properties, centered on the first protostellar core to form. The quantities are mass-weighted and taken from a slice through the midplane of the disk. In the lower right-hand plot we show the radial distribution of the disk's Toomre parameter, $Q = c_{\rm s} \kappa / \pi G \Sigma$, where $c_{\rm s}$ is the sound speed and $\kappa$ is the epicyclic frequency. Beause our disk is Keplerian, we adopted the standard simplification, and replaced $\kappa$ with the orbital frequency. The molecular fraction is defined as the number density of hydrogen molecules ($n_{\rm H_{2}}$), divided by the number density of hydrogen nuclei ($n$), such that fully molecular gas has a value of 0.5}
\label{pcc_fig2}
\end{figure}

After around 90 years, the disk had nearly doubled in size. For the first 60 years, the structure of the disk was dominated by a strong two-arm spiral pattern, a feature common to simulations of present-day star formation \cite{Krumholz2009}. Spiral structures of this kind are a signature of self-gravitating disks, in which gravitational torque provides the main source of angular momentum transport. Although the spiral pattern started out fairly symmetric, it quickly developed non-axisymmetric features. Eventually, the gas in one of the spiral arms became locally unstable and started to collapse, forming a second protostar at a distance of roughly 20 AU from the primary. The mass of the central protostar at this point was only 0.5 \solmas.

The surface density remained roughly constant as the disk grew, with the temperature behaving in a similar fashion and thus the ability of the disk to transport angular momentum remained the same as the disk grew. This can be seen by considering the Toomre `$Q$' parameter (Fig. ~\ref{pcc_fig2}), which provides a measure of the gravitational instability of the disk. For high $Q$, the disk is stable, while for values around 1, the disk is formally unstable to fragmentation. As the disk  grew, the value of $Q$ remained around 1 in the outer regions, and so the dynamics of the disk were dominated by gravitational instabilities.
  
\begin{figure}
\centerline{
	\includegraphics[height=5.in]{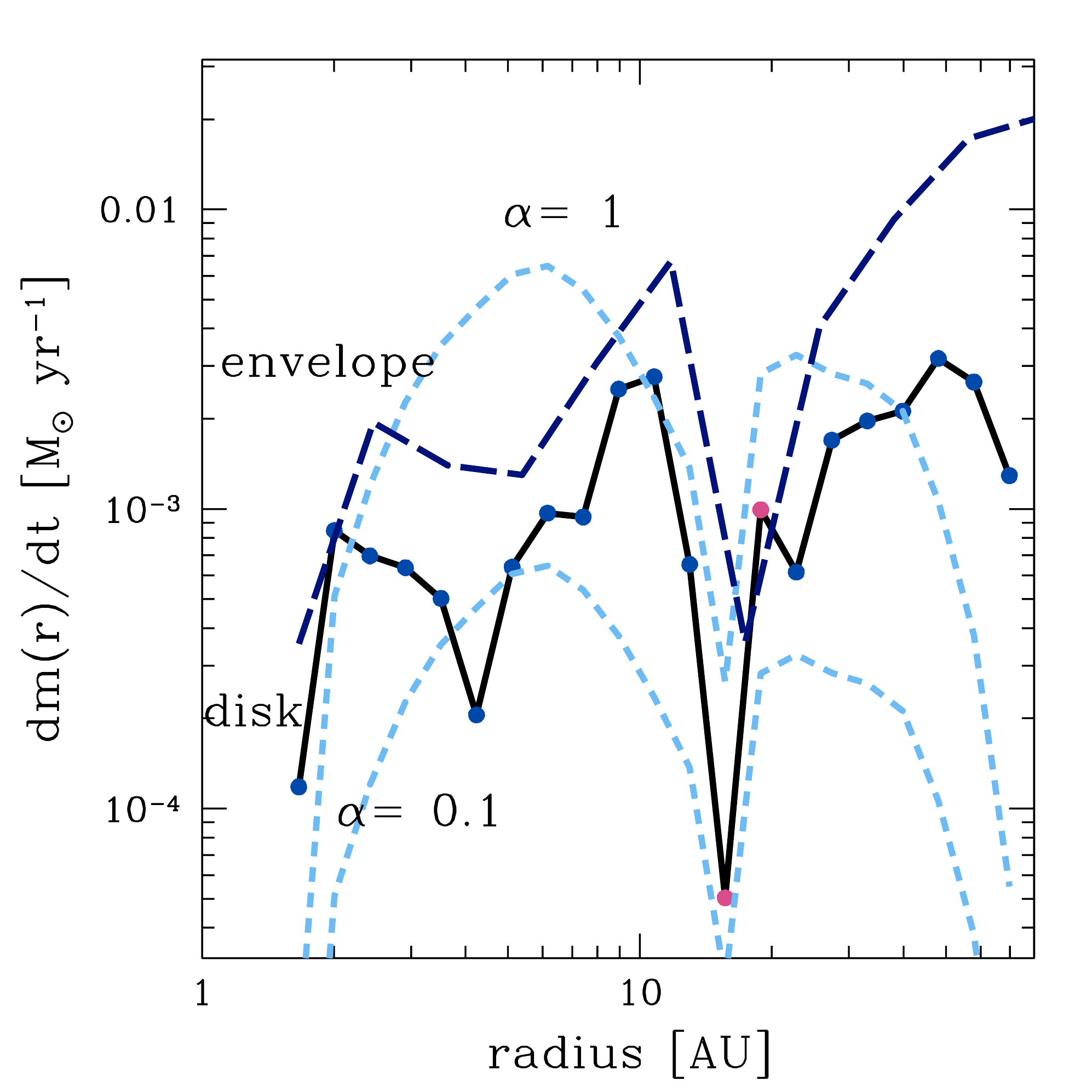}
}
\caption{The mass transfer rate through the disk is denoted by the solid black line, while the mass infall rate through spherical shells with the specified radius is shown by the dark blue dashed line. The latter represents the total amount of material flowing through a given radius, and is thus a measure of the material flowing through {\em and onto} the disk at each radius. Both are shown at the onset of disk fragmentation.  In the case of the disk accretion we have denoted annuli that are moving towards the protostar with blue dots, and those moving away in pink (further details can be found in Section 6 of the online material). The light blue dashed lines show the accretion rates expected from an `alpha' (thin) disk model, where $\dot{M}(r) = 3\,\pi\,\alpha\,c_{\rm s} (r)\,\Sigma(r)\,H(r)$, with two global values of alpha and where $c_{\rm s}(r)$, $\Sigma(r)$, and $H(r)$ are (respectively) the sound speed, surface density and disk thickness at radius $r$.}
\label{pcc_fig3}
\end{figure}
 
Figure~\ref{pcc_fig3}, which shows the accretion rate through the disk and  envelope as a function of radius from the central protostar, helps to explain why the disk became so unstable. The accretion rate through the disk was considerably lower than the rate at which material was added to the disk. If one assumes a simple $\alpha$-disk model \cite{ss1973}, then the disk accreted with an effective $\alpha$ between 0.1 and 1, which is typical for gravitational torques in strongly self-gravitating disks \cite{lr2005}. Although this is an efficient mechanism for transporting angular momentum, it was nevertheless unable to process the material in the disk quickly enough before more was added from the infalling envelope. As a result, the disk grew in mass, became gravitationally unstable, and ultimately fragmented. 

For a region of the disk to form a new star, it must be able to rid itself of the heat generated as it collapses, and further, it must do so before the motions in the disk shear the region apart \cite{gammie01}. In the case of a primordial protostellar disk, this cooling is provided by molecular hydrogen. At typical disk number densities, around 10$^{12}$ to 10$^{14}$ cm$^{-3}$, most of the cooling comes from H$_{2}$ line emission. In the centers of the spiral arms, or at number densities of around $10^{14}$ cm$^{-3}$, cooling by collision-induced emission (CIE) from H$_{2}$ begins to dominate, and it is this process that allowed the initial gravitational fragmentation to take place in the spiral arms (see Section 6 of $10$).  At number densities of around $10^{16}$ cm$^{-3}$, the gas becomes optically thick to CIE, and begins to heat up further as it is compressed. However, it turns out that primordial protostellar accretion disks are unusual in that the dissociation of molecular hydrogen plays an important role in regulating their temperature evolution. In contrast to present-day star formation, primordial protostellar disks are remarkably hot (Fig.~\ref{pcc_fig2}). Any further compression causes the hydrogen molecules to dissociate, removing energy from the system. This process acts like a thermostat, keeping the temperature in the collapsing region roughly constant while the collapse proceeds. Thus, perhaps somewhat counter-intuitively, it is the relatively high temperature associated with primordial star formation, coupled with a high molecular fraction, that allowed the protostellar accretion disk to fragment.

Fragmentation does not stop with the formation of a second protostar. Only four years after the initial fragmentation of the disk, the dense ridge in a neighboring spiral arm also fragmented, forming a third protostar. Fifteen years later, the disk fragmented yet again and the protostars evolved into a chaotic multiple system. At the stage of the evolution shown in Fig.~\ref{pcc_fig1}, slightly less than 1 ${\rm M_{\odot}}$ of gas had been converted into protostars, and hence the system was still very much in its infancy. Eventually, beyond the epoch followed by our current calculation, the disk will lose its ability to fragment once the heating from the protostars becomes strong enough to dissociate all of the H$_{2}$, thereby removing the major coolant \cite{tm2004}. However the system will continue to accrete infalling gas from the collapsing envelope, until one of the protostars becomes sufficiently hot to ionize the gas, which finally terminates the accretion \cite{mt2008}.

Are our results representative of primordial star formation in general, or is the gas within our halo special in some way? The properties of the gas that assembled the disk in our simulation were remarkably similar to those found in other simulations of primordial star formation \cite{SOM}. Crucially, the angular momentum of the inner 4 ${\rm M_{\odot}}$ -- the combined mass of the disk and the protostar at the point of fragmentation -- was $1.4 \times 10^{-3}$ km\,s$^{-1}$\,pc, comparable to the $1.2 \times 10^{-3}$ km\,s$^{-1}$\,pc reported elsewhere \cite{abn2002}. Fragmentation of the kind we have seen should therefore be a normal part of the formation of primordial stars, suggesting that the endpoint of the Pop III star formation process is significantly more complicated than previously thought. Idealized calculations have previously found hints of such widespread binarity among the first stars \cite{machida08}, but our work establishes this property for realistic cosmological initial conditions, coupled with the radiative feedback, and the extremely high resolution needed to address protostellar disk evolution.

While our simulations do not show how primordial disks evolve beyond the initial disk fragmentation. present-day star formation calculations \cite{Krumholz2009, peters2010a, peters2010b} predict that as fragmentation proceeds in high-mass accretion disks, new protostars form at increasingly larger radii. Because gas can only be accreted when its angular momentum matches that of the protostar, it is easier for the new objects than for the preexisting ones to gain fresh gas that moves inwards through the accretion disk, tending to drive the system towards equal masses \cite{bb1997}. If fragmenting Pop III systems evolved in a similar fashion, then one plausible outcome of primordial star formation would be the formation of nearly equal-mass Pop III binaries. Their potential existence would strengthen the case for high-redshift gamma-ray bursts origination from the first stars\cite{bl06}. From present-day star formation, we also know that young multiple systems are dynamically unstable, over time leading to the dynamical ejection of protostars \cite{pmg2010} with the preferentially close, high-mass binary systems remaining in the center. If Pop III stars were dynamically ejected from such systems before accreting very much gas, then there is the possibility that some of these stars may have had masses low enough for them to have survived until the present-day.

\nocite{thanks}
\bibliography{primdisc}

\begin{thebibliography}{10}
\bibitem{SOM}
Materials and methods are available as supporting material on Science Online.
\end{thebibliography}

\bibliographystyle{Science}

% Following is a new environment, {scilastnote}, that's defined in the
% preamble and that allows authors to add a reference at the end of the
% list that's not signaled in the text; such references are used in
% *Science* for acknowledgments of funding, help, etc.

%\begin{scilastnote}

\clearpage

%
%
%
%
%
% %%%%%%%%% THE SOM....
%
%
%
%
%
%
%

\newpage

\vspace{6cm}
\begin{center}
\Large{Supporting Online Material for The Formation and Fragmentation of Disks around Primordial Protostars} \\
\vspace{3cm}
\large{Paul~C.~Clark,$^{\ast}$ Simon~C.~O. Glover, Rowan~J. Smith, Thomas~H. Greif,
Ralf S. Klessen, Volker Bromm} \\
\normalsize{$^\ast$To whom correspondence should be addressed; E-mail:  pcc@ita.uni-heidelberg.de}
\end{center}

\newpage

\section{Numerical methods}
\label{method}
The simulations discussed in this paper were performed using a modified version
of the Gadget 2 smoothed particle hydrodynamics (SPH) code \cite{Springel05}.
Our modifications involve the addition of a treatment of primordial chemistry and
cooling, along with a scheme for replacing gas in unresolved regions with 
collisionless ``sink'' particles \cite{BBP95,JAP05}.  We have also implemented
an approximate model for the effects of the accretion luminosity produced by the 
accreting protostars. Details of these modifications are given  below.

\paragraph{Chemistry and cooling}
To model the chemical evolution of the metal-free gas, we use a chemical network
consisting of 45 reactions amongst twelve chemical species: H, H$^{+}$, H$^{-}$, 
H$_{2}^{+}$, H$_{2}$, He, He$^{+}$, He$^{++}$, D, D$^{+}$, HD, and free electrons
\cite{clark10}. For most of these reactions, we use the same reaction rate coefficients as in 
Clark et~al.\ \cite{clark10}. The exceptions are the three-body reaction
\begin{equation}
{\rm H} + {\rm H} + {\rm H} \rightarrow {\rm H_{2}} + {\rm H}, \label{tb1}
\end{equation}
and its inverse reaction, the collisional dissociation of H$_{2}$ by H,
\begin{equation}
{\rm H_{2}} + {\rm H}  \rightarrow  {\rm H} + {\rm H} + {\rm H}. \label{tb2}
\end{equation}
Considerable uncertainty exists concerning the rate of reaction~\ref{tb1} at
low temperatures \cite{glover08}. We adopt the rate coefficient given for this reaction
by Abel et~al.~\cite{abn02}, which is the smallest of those in common usage, and
hence will yield the highest temperature for the dense gas in the center of the
minihalo \cite{GS09}. This warmer gas is less likely to fragment than the cooler gas
yielded by alternative choices for the three-body reaction rate coefficient. Moreover,
accretion disks formed in simulations run with a small value for the three-body rate
coefficient are less likely to be H$_{2}$-dominated than those formed in simulations
using a larger value for the rate coefficient, and so again will be more stable against
fragmentation. In view of this, we consider our choice of the Abel et~al.\ rate coefficient 
to be conservative, in the sense that it will tend to reduce the likelihood of fragmentation.

Our decision to use a different rate coefficient for Reaction~\ref{tb1} than in the Clark
et~al.\ study means that we must also adopt a different rate coefficient for 
reaction~\ref{tb2}, as at high densities the ratio between the rate coefficients of 
reactions 1 and 2 must satisfy
\begin{equation} 
\frac{k_{1}}{k_{2}} = K,
\end{equation}
where $k_{1}$ and $k_{2}$ are the rate coefficients for reactions \ref{tb1} and \ref{tb2},
respectively, and where $K$ is the equilibrium constant for the pair of reactions, 
given by \cite{fh07}
\begin{equation}
K = 1.05 \times 10^{-22} T^{-0.515} \exp \left(\frac{52000}{T} \right). \label{eq_const}
\end{equation}
Note that the value of $K$ does not depend on our choice of values for $k_{1}$ or 
$k_{2}$, and hence we cannot vary these rate coefficients independently: if we change
$k_{1}$, we must also adjust $k_{2}$ such that Equation~\ref{eq_const} remains satisfied.

Our treatment of the cooling also follows the Clark et~al.\ study. We account for the full
set of processes described in that paper, including electronic excitation of H, He 
and He$^{+}$, cooling from the recombination of H$^{+}$ and He$^{+}$, 
Compton cooling and bremsstrahlung. However, in practice only a few processes play 
an important role at the densities and temperatures characteristic of the protostellar 
accretion disk -- H$_{2}$ rotational and vibrational line cooling, H$_{2}$ 
collision-induced emission (CIE) cooling, H$_{2}$ collisional dissociation cooling, 
and heating due to three-body H$_{2}$ formation. 

We model H$_{2}$ line cooling in the optically thin regime using a comprehensive treatment
that includes the effects of collisions between H$_{2}$ molecules and H and He atoms, other H$_{2}$ 
molecules, protons and electrons \cite{ga08}, and that accounts for the
transition to local thermodynamic equilibrium level populations at number densities 
$n \gg 10^{4} \: {\rm cm^{-3}}$. At densities $n \sim 10^{9} \: {\rm cm^{-3}}$ and above, 
the strongest of the H$_{2}$ ro-vibrational lines become optically thick, reducing the 
effectiveness of H$_{2}$ line cooling. To account for this effect, we use an approach
based on the Sobolev approximation \cite{yoha06}.  We write the H$_{2}$ cooling rate  as
\begin{equation}
\Lambda_{\rm H_{2}} = \sum_{\rm u,l} \Delta E_{\rm ul} A_{\rm ul} \beta_{\rm esc, ul} n_{\rm u},
\end{equation}
where $n_{\rm u}$ is the number density of hydrogen molecules in upper energy
level $u$, $\Delta E_{\rm ul}$ is the energy difference between this upper level
and a lower level $l$, $A_{\rm ul}$ is the spontaneous radiative transition rate
for transitions between $u$ and $l$, and $\beta_{\rm esc, ul}$ is the escape
probability associated with this transition, i.e.\ the probability that the emitted photon 
can escape from the region of interest. We fix $n_{\rm u}$ by assuming that the 
H$_{2}$ level populations are in local thermodynamic equilibrium, and write the
escape probabilities for the various transitions as \cite{yoha06}
\begin{equation}
\beta_{\rm esc, ul} = \frac{1 - \exp(-\tau_{\rm ul})}{\tau_{\rm ul}},
\end{equation}
where we use the approximation that
\begin{equation}
\tau_{\rm ul} \simeq \alpha_{\rm ul} L_{\rm s} \label{tau}
\end{equation}
where $\alpha_{\rm ul}$ is the line absorption coefficient and $L_{\rm s}$ is the
Sobolev length. In the classical, one-dimensional spherically symmetric case,
the Sobolev length is given by
\begin{equation}
L_{\rm s} = \frac{v_{\rm th}}{|{\rm d}v_{\rm r} / {\rm d}r|},
\end{equation}
where $v_{\rm th}$ is the thermal velocity, and ${\rm d}v_{\rm r} / {\rm d}r$ is the
radial velocity gradient. In our three-dimensional simulations, we generalize 
this as \cite{nk93}
\begin{equation}
L_{\rm s} = \frac{v_{\rm th}}{|\nabla \cdot {\mathbf v}|},
\end{equation}
To prevent the H$_{2}$ cooling rate from being reduced by an unphysically large 
amount in regions with small velocity gradients, we limit $L_{\rm s}$ to be less than
or equal to the local Jeans length, $L_{\rm J}$. We justify this choice of limit by noting 
that there are strong density gradients in the gas on scales $L \gg L_{\rm J}$, and so
we expect the bulk of the contribution to the H$_{2}$ line absorption to come from material 
within only a few Jeans lengths. 

\begin{figure}
\centerline{
\includegraphics[scale=0.55]{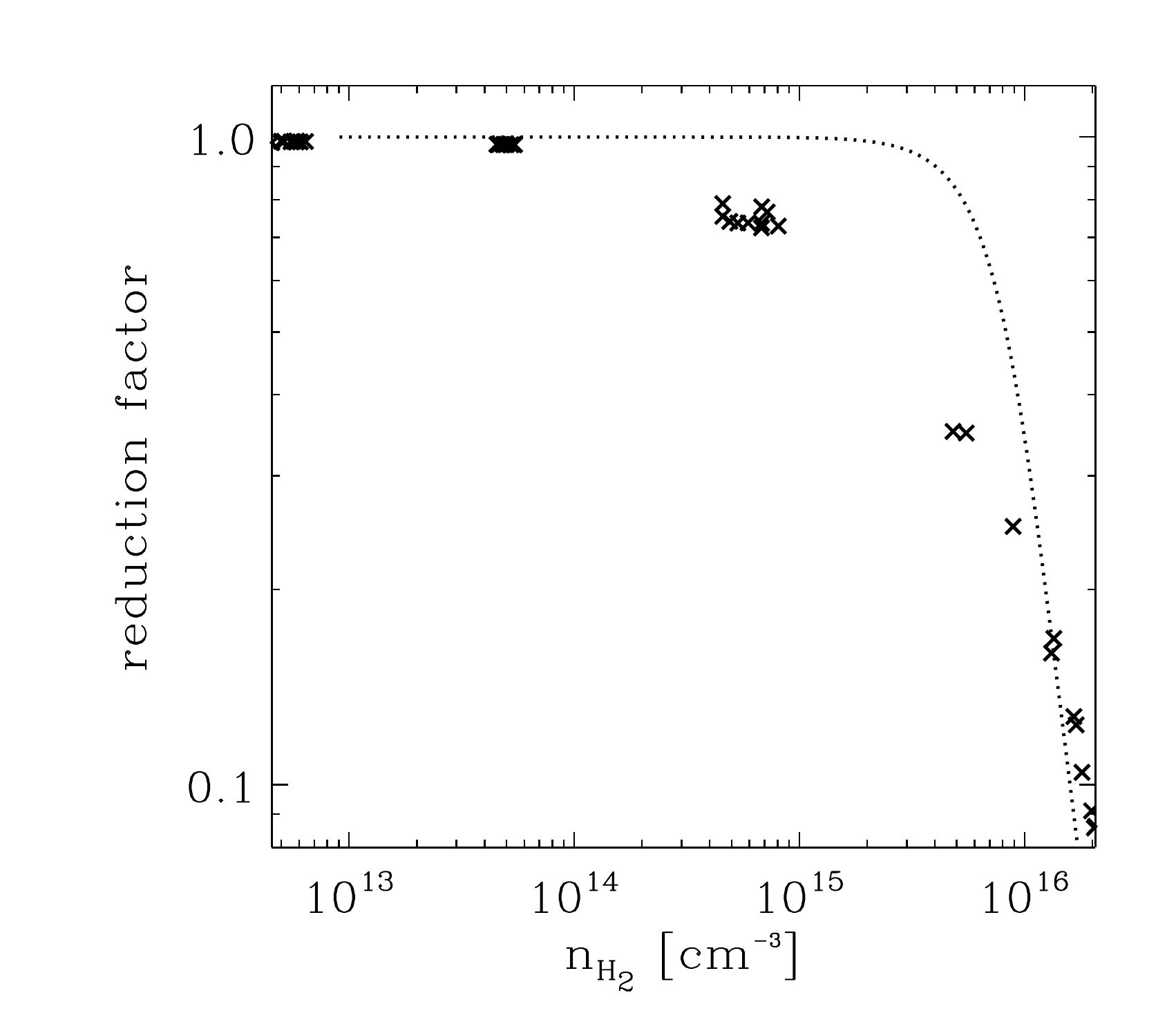}
}
\caption{\label{cie} The ratio $\Lambda_{\rm CIE, thick} / \Lambda_{\rm CIE, thin}$, plotted as a function of the
number density of H$_{2}$. The solid line gives the values predicted by our empirical prescription 
(Eqs.~\ref{eq:reduction}--\ref{eq:tau_cie}), while the individual data points show the results of the more 
sophisticated procedure described in the text.}
\end{figure}

At very high densities ($n > 10^{14} \: {\rm cm^{-3}}$), CIE cooling from H$_{2}$ becomes
more effective than H$_{2}$ line cooling \cite{ra04}. We model this as discussed in 
Clark et al., and account for the reduction of the CIE cooling rate due to the effects of
continuum opacity using the following empirical prescription, based on previous 
high-resolution 1D calculations of the formation of primordial protostars \cite{ra04,r02}
\begin{equation}
\Lambda_{\rm CIE, thick} = \Lambda_{\rm CIE, thin} {\rm min}\left(\frac{1 - 
e^{-\tau_{\rm CIE}}}{\tau_{\rm CIE}}, 1 \right),
\label{eq:reduction}
\end{equation}
where 
\begin{equation}
\tau_{\rm CIE} = \left(\frac{n_{\rm H_{2}}}{7 \times 10^{15} \: {\rm cm^{-3}}} \right)^{2.8}.
\label{eq:tau_cie}
\end{equation}
The reduction in the CIE cooling rate predicted by this prescription agrees to within a
factor of a few with that measured in a more recent 3D simulation of Pop.\ III star formation
\cite{yoh08}, but this simulation followed the evolution of the gas only prior to the formation
of the initial protostar, and hence could not test the validity of the empirical prescription at later
times in the evolution of the system. In particular, it is unclear whether Eq.~\ref{eq:tau_cie} 
remains a good approximation once material begins to build up in the protostellar accretion disk. 
To test this, we selected a set of SPH particles covering a range of different number densities
in the output snapshot produced immediately prior to the formation of the second sink particle.
We computed continuum opacities along forty-eight different lines of sight surrounding each SPH 
particle by solving the equation
\begin{equation}
 \tau_{\rm i} = \int_{0}^{L_{i}} \rho(l) \kappa_{\rm P}(\rho, T, x_{\rm H_{2}}) {\rm d}l
\end{equation}
along each line of sight, where $L_{i}$ is the distance from the particle to the edge of the simulation
volume, $\rho(l)$ is the local density at a distance $l$ along the line of sight, $T$ is the temperature
at the same point, $x_{\rm H_{2}} = n_{\rm H_{2}}/n$, and $\kappa_{\rm P}$ is the Planck mean opacity 
of the gas. The mean opacity is a function of the local density, temperature and chemical composition, and
we calculated it by interpolation from previously tabulated values \cite{md05}. An estimate of $\tau_{\rm CIE}$
then follows from
\begin{equation}
\exp(-\tau_{\rm CIE}) = \frac{1}{48} \sum_{i=1}^{48} \exp(-\tau_{i}).
\end{equation}
We then used Eq.~\ref{eq:reduction} together with this estimate of the CIE opacity to compute the reduction 
in the CIE cooling rate, and compared this with the reduction predicted by our empirical prescription.  The
results are plotted in Figure~S\ref{cie}.  We see that in practice, the continuum opacity begins to reduce the
CIE cooling rate somewhat earlier than would be predicted by Eqs.~\ref{eq:reduction}--\ref{eq:tau_cie}, 
but that at densities above $n \sim 10^{16} \: {\rm cm^{-3}}$, the two different techniques yield very similar
answers. Moreover, the difference between the reduction factor (i.e.\ the ratio of the optically thick and optically
thin CIE cooling rates) predicted by Eqs.~\ref{eq:reduction}--\ref{eq:tau_cie} and the one computed from our
simulation results is never greater than a factor of two over the whole of the regime where CIE cooling is 
important. Given the very strong temperature dependence of the CIE cooling ($\Lambda_{\rm CIE} \propto
T^{4}$), a factor of two uncertainty in the cooling rate translates into only a 20\% uncertainty in the gas 
temperature, which is unlikely to be large enough to significantly affect the conclusions of our study. 
Furthermore, it should also be noted that gas at the typical densities of $n \sim 10^{13}$--$10^{15} \: {\rm cm^{-3}}$ 
found in the accretion disk remains in the optically thin regime as far as CIE cooling is concerned. The effects of 
continuum opacity only become important in gas that is already undergoing run-away gravitational collapse.

\begin{figure}
\centerline{
\includegraphics[scale=0.55]{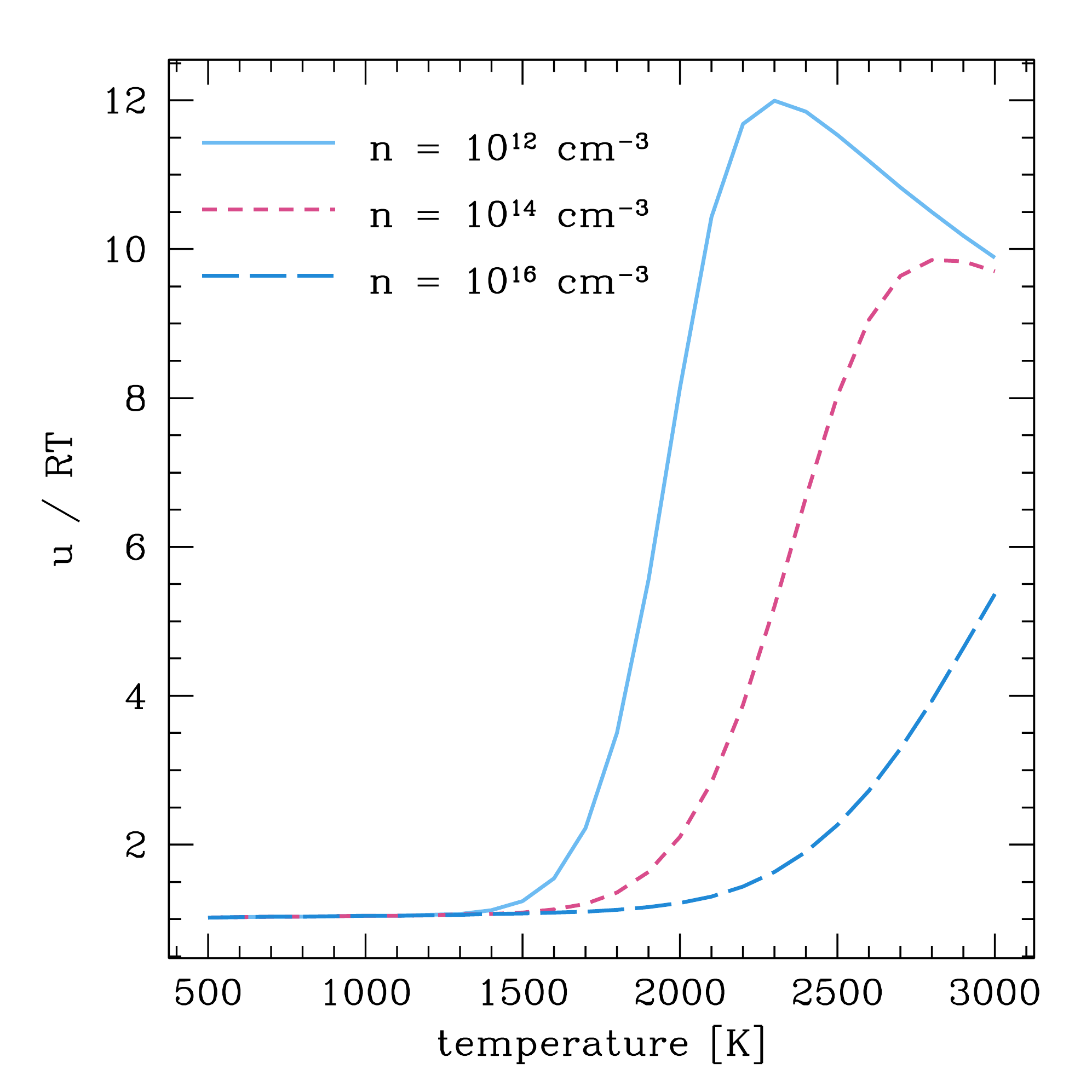}
}
\caption{The specific heat capacity $u / RT$ of primordial gas (where $u$ is the specific energy and 
$R = k_{\rm B} / m_{\rm H}$), computed assuming chemical equilibrium, and plotted as a function of temperature 
for three different values of the hydrogen nuclei number density $n$. At low temperatures, the specific heat capacity
is almost independent of $T$, but it increases sharply above $T \sim 1500$--2000~K as the H$_{2}$ in the gas 
begins to dissociate.  \label{fig:eos}}
\end{figure}

Finally, our thermal model also accounts for changes in the internal energy of the gas due to the dissociation 
or formation of H$_{2}$ molecules. Every collisional dissociation of an H$_{2}$ molecule requires 4.48~eV
of energy (the binding energy of the H$_{2}$ molecule), which is removed from the thermal energy of the gas,
while every time that a new H$_{2}$ molecule is formed by the three-body process, the gas gains 4.48~eV
of thermal energy. For a gas in chemical equilibrium, the simplest way to account for these effects is through
the equation of state of the gas. Given the hydrogen nuclei number density $n$ and the gas temperature $T$,
the equilibrium molecular fraction $x_{\rm H_{2}} = n_{\rm H_{2}} / n$ can be determined from the Saha equation, 
and the specific internal energy $u$ can then be obtained by summing up the contributions due to the translational
motions of the hydrogen molecules, hydrogen atoms and helium atoms, the internal degrees of freedom of the
hydrogen molecules, and the dissociation energy of the H$_{2}$ molecules \cite{bb75}. A convenient way to 
visualize how $u$ varies with temperature is to examine the behaviour of the specific heat capacity, $u / RT$, 
where $R = k_{\rm B} / m_{\rm H}$, $k_{\rm B}$ is Boltzmann's constant, and $m_{\rm H}$ is the mass of a hydrogen
atom. In Figure~S\ref{fig:eos}, we show how $u / RT$ varies with temperature for several different values of $n$.
At low temperatures, the specific heat capacity is almost independent of temperature, but above a temperature 
of $1500$--2000~K, it rises sharply as the H$_{2}$ begins to dissociate. Physically, this means that once the gas
reaches this temperature regime, a much greater amount of energy is required in order to raise the temperature
further than would be the case in colder gas. We see this effect in our simulations, in the gas undergoing run-away
gravitational collapse in the disk: despite being strongly heated by compression, it increases its temperature only
slowly once it reaches the temperature regime in which H$_{2}$ dissociation begins to occur.

Unfortunately, although this approach is useful for providing insight into the thermodynamical behaviour of the gas,
it relies on the gas being in chemical equilibrium, which is not the case for the majority of the gas in our simulations.
Chemical equilibrium is reached only for gas number densities above $n \sim 10^{15} \: {\rm cm^{-3}}$  \cite{yoh08}.
Therefore, rather than account for H$_{2}$ formation heating and dissociation cooling by including their effects in
our adopted equation of state, we instead follow previous studies \cite{yoha06,yoh08} and include a chemical
heating/cooling term $\dot{e}_{\rm chem} = \chi_{\rm H_{2}} \dot{n}_{\rm H_{2}}$ in our equation for the time evolution
of the internal energy density of the gas, where $\chi_{\rm H_{2}} = 4.48 \: {\rm eV}$ and $\dot{n}_{\rm H_{2}}$ is the
rate of change of the H$_{2}$ number density, computed with the mass density $\rho$ held constant.

\paragraph{Sink particles}
We create sink particles following the standard SPH prescription \cite{BBP95}, and 
adopt a threshold density for sink particle creation $n_{\rm th} = 10^{17} \: {\rm cm^{-3}}$.
At this density, radiative cooling of the gas has become almost completely ineffective,
and the dominant cooling process is collisional dissociation of H$_{2}$ (see Section~\ref{sec:thermo}
below). We therefore do not expect significant fragmentation of the gas to occur at
densities $n > n_{\rm th}$. Any SPH particle that exceeds our threshold density
is potentially eligible to be converted into 
a sink particle, but sink particle creation occurs only for particles satisfying several additional
criteria. First, the candidate sink particle must be located a distance $r \geq 2 r_{\rm acc}$
away from any other sink particle, where $r_{\rm acc}$ is the sink particle accretion radius
(see below). This criterion prevents sink particle formation from occurring within gas that
is about to be accreted by another sink particle. Second, we check to see whether the 
smoothing length of the particle is less than the accretion radius of the sink particle that 
it will become. This test ensures that we are resolving the structure of
the gas within the sink particle accretion radius. Third, we ensure that the candidate
sink particle and its nearest neighbors are all on the same integration time-step. 
Once these preliminary criteria are met, the dynamical state of the candidate sink
particle and its neighbors is examined. In order to ensure that sink particle creation
occurs only within gas which is undergoing gravitational collapse, we require that
the total energy of the candidate sink plus the gas within one smoothing length be
negative, implying that the collection of SPH particles are gravitationally bound, and
also verify that the divergence of the particle accelerations is negative. This final check 
ensures that the group of particles is not in the process of being tidally disrupted or
bouncing. If all of these tests are passed, the candidate particle is converted into a sink
particle. Its fifty nearest neighbors are removed from the simulation, and their masses 
and linear momenta are added to the sink particle.

As the simulation progresses, sink particles are allowed to accrete other gas particles
that move to within $r < r_{\rm acc}$ of the sink particle. However,
gas particles within the accretion radius are accreted only if they pass several additional
tests. First, they must be gravitationally bound to the sink particle, and moving towards it.
Second, if more than one sink is present, they must be bound more strongly to the candidate
sink than to any other sink in the simulation. Third, the gas particle and the sink must be on 
the same integration time-step. Once these tests are passed, the mass and linear momentum 
of the SPH particle are added to the sink particle, and the SPH particle is removed from the
simulation. 

In the simulations presented in this paper, we set $r_{\rm acc} = 1.5$~AU. Gravitational
forces are softened on a significantly smaller length scale of 0.066~AU to ensure that the 
force exerted by the sink particle on gas outside of the accretion radius is essentially the
same as that of a point mass, and to permit sinks to come close enough together to allow
us to check for potential mergers. For comparison, the minimum SPH smoothing length
resolvable in the simulation is 0.0551~AU, and so we resolve pressure gradients on a similar
scale to gravitational forces, as is required to prevent artificial fragmentation \cite{bb97}.
We check for mergers between sinks by examining whether any 
two sinks ever come within two protostellar radii of each other, taking $50 \: R_{\odot}$
as a reasonable estimate for the pre-main sequence radii of the Population III stars
studied in this simulation (see Eq.~\ref{eq:rstar} below). In this study, we find no evidence
for any potential mergers.

We have also performed a simulation with a larger sink accretion radius of 3~AU. 
The gravitational softening length in this simulation was again fixed at 0.066~AU.
This simulation produced qualitatively similar results to those from an $r_{\rm acc} = 1.5$~AU 
simulation with the same assumed protostellar accretion rate. In the simulation with the
larger accretion radius, the first signs of fragmentation appeared at a distance of 8~AU from 
the central protostar, roughly twice the separation seen in the simulation with $r_{\rm acc} = 
1.5$~AU. As in the simulations with smaller $r_{\rm acc}$, the disk proceeded to fragment
further.  The difference between the two simulations is easily explained. When the sink particles 
have smaller accretion radii, we are able to resolve fragmentation that occurs on 
smaller scales. In this particular case, the simulation performed using an accretion radius of
1.5~AU was able to model the formation of an object that was swallowed by the sink 
particle in the simulation performed with the 3~AU accretion radius. The third object to form in
the $r_{\rm acc} =  1.5$~AU simulation is located in roughly the same place as the second
object in the $r_{\rm acc} =  3$~AU simulation, demonstrating that the evolution of the disk
at distances $r \gg r_{\rm acc}$ is largely unaffected by the choice of $r_{\rm acc}$.

\paragraph{Accretion luminosity}
We treat each sink particle formed in our simulation as a separate protostar, and
account for the energy released by accretion onto the surfaces of these protostars.
To model the effects of this accretion luminosity, we first compute the bolometric
accretion luminosity for each accreting protostar
\begin{equation}
 L_{\rm acc} = \frac{G \dot{M}_{*} M_{*}}{R_{*}}, \label{eq:lacc}
\end{equation}
where $\dot{M}_{*}$ is the accretion rate onto the protostar, $M_{*}$ is the
protostellar mass and $R_{*}$ is the protostellar radius. In this study,
we model only the first few hundred years of evolution of the protostellar
system. Since the Kelvin-Helmholtz relaxation timescale of even a high-mass
protostar is of the order of a few thousand years \cite{ho09}, 
we can be confident that the
protostars in our simulation will not yet have thermally relaxed, and will 
still be in the initial adiabatic accretion phase. We therefore relate the
protostellar radius to the current protostellar mass and mass accretion rate 
using the following relationship \cite{sps86}
\begin{equation}
R_{*} = 26 R_{\odot} \left( \frac{M_{*}}{{\rm M}_{\odot}} \right)^{0.27} 
\left(\frac{\dot{M}_{*}}{10^{-3} \: {\rm M_{\odot} \: yr^{-1}}} \right)^{0.41},
\label{eq:rstar}
\end{equation}
which was derived for adiabatically accreting, metal-free protostars embedded in a 
spherically symmetric inflow. More recent calculations for the case of rotating infall 
\cite{tm04} find an even larger value for $R_{*}$ during the adiabatic accretion phase
(see their Fig.~5), and so using the expression given by Equation~\ref{eq:rstar}
provides us with a conservative upper limit on the bolometric accretion luminosity during 
this phase.

Combining Eqs.~\ref{eq:lacc} and \ref{eq:rstar}, we find that
\begin{equation}
 L_{\rm acc} \simeq 1200 {\rm L}_{\odot}  \left( \frac{M_{*}}{{\rm M}_{\odot}} \right)^{0.73} 
\left(\frac{\dot{M}_{*}}{10^{-3} \: {\rm M_{\odot} \: yr^{-1}}} \right)^{0.59}.
\label{eq:lacc2}
\end{equation}
Using this expression for the bolometric accretion luminosity, we next determine
the heating rate of the gas surrounding the protostar from
\begin{equation}
\Gamma_{*} = \rho \kappa_{\rm P} \frac{L_{\rm acc}}{4 \pi r^{2}},
\label{eq:heat}
\end{equation}
where $\rho$ is the mass density, $r$ is the distance to the protostar,
and $\kappa_{\rm P}$ is the Planck mean opacity of the gas. We calculate
this mean opacity by interpolation, using tabulated values that include the effects 
of both line and continuum absorption, and that account for the extra electrons provided
by ionized lithium \cite{md05}. Equation~\ref{eq:heat} assumes that the gas is optically 
thin to the radiation from the accreting protostar. In reality, this is unlikely to be
the case. However, making this assumption allows us to avoid the extremely high 
computational cost that would be associated with an accurate treatment of
the transfer of the protostellar radiation, and also gives us a conservative
upper limit on the effectiveness of the protostellar feedback.  

\begin{figure}
\centerline{
\includegraphics[scale=0.5]{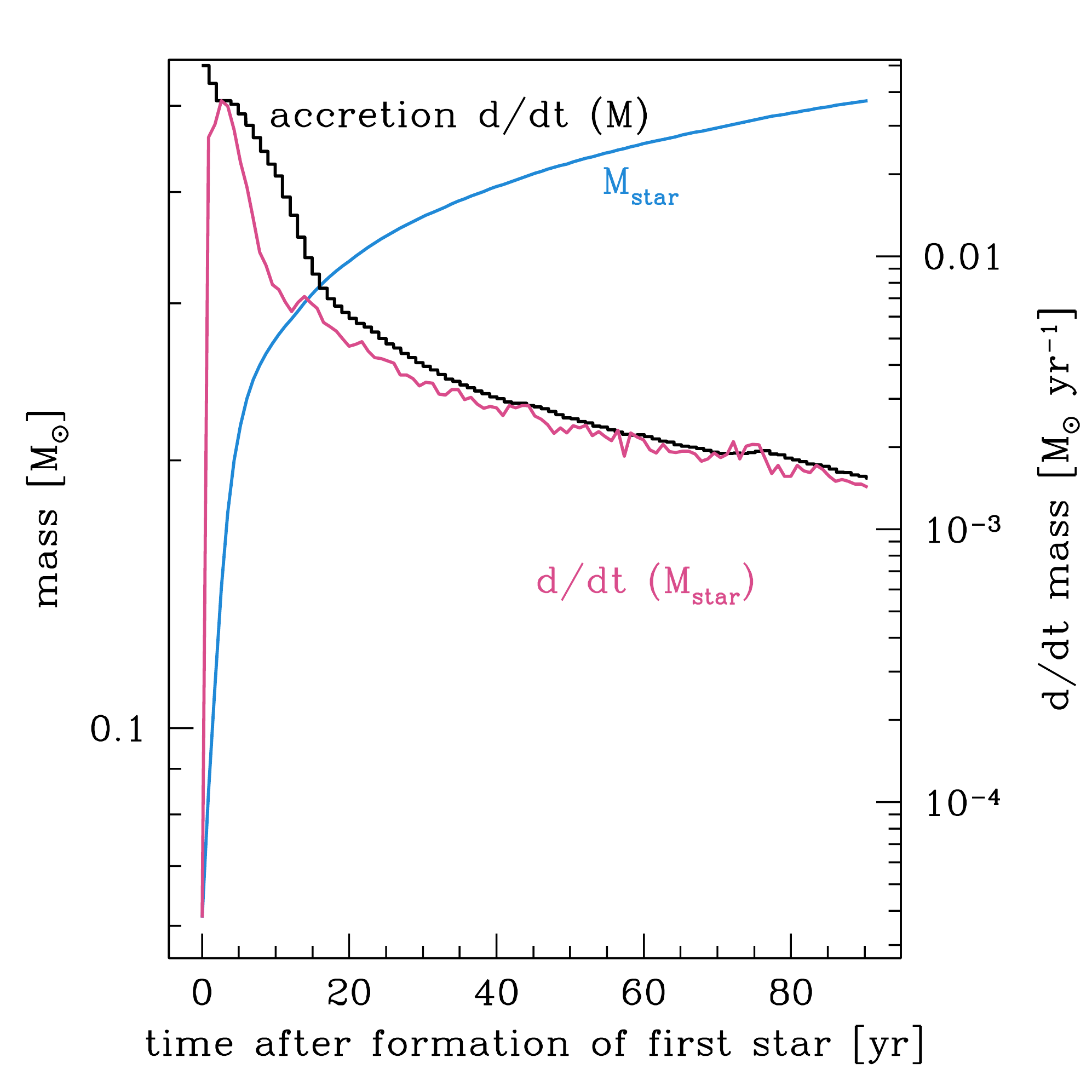}
}
\caption{Evolution with time of the mass of the first sink particle ({\it blue curve}),
the true accretion rate onto the protostar ({\it magenta curve}) and the time-averaged
estimate of the accretion rate used to computing the accretion luminosity ({\it
black curve}). Our estimate is a slight overestimate of the true accretion rate, particularly
at early times when the accretion rate is changing rapidly with time.
\label{accr_rate}}
\end{figure}

\begin{figure}
\centerline{
\includegraphics[scale=0.39]{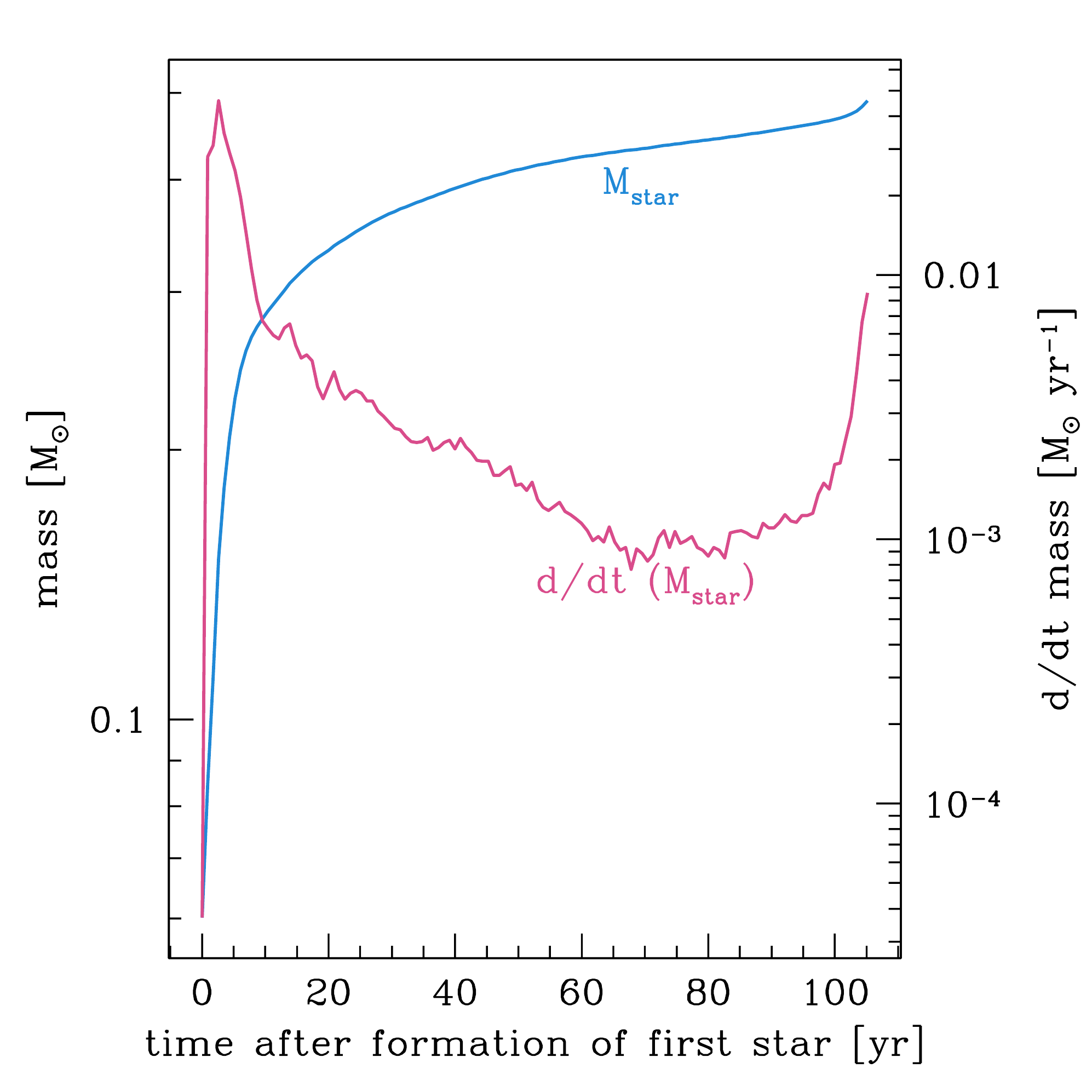}
\includegraphics[scale=0.39]{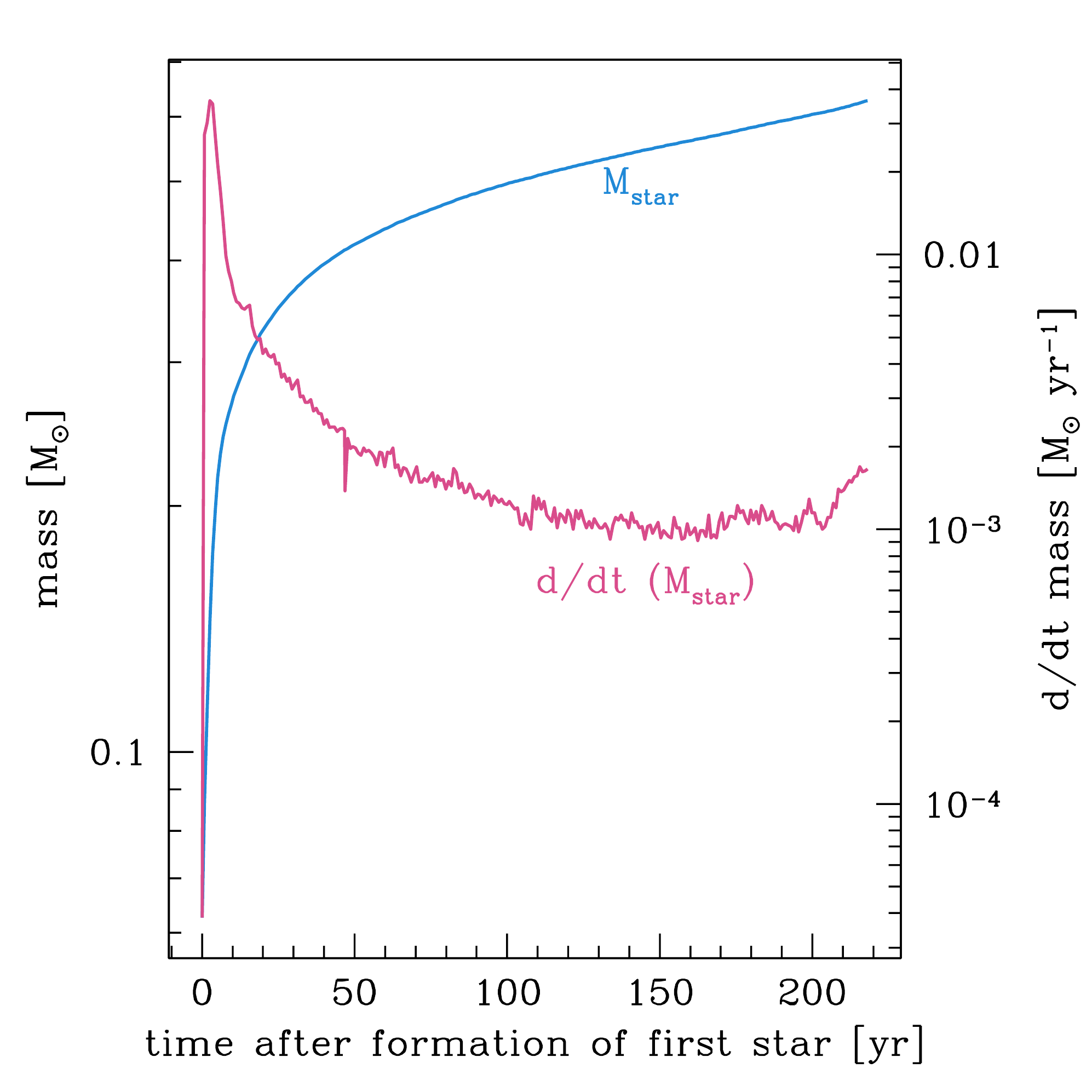}
}
\caption{{\it Left panel:} Evolution with time of the mass of the first 
sink particle ({\it blue curve}) and the accretion rate
onto that sink particle ({\it magenta curve}) in the simulation in which
we assume an accretion rate of $10^{-3} \: {\rm M_{\odot} \: yr^{-1}}$
for the purposes of computing the accretion luminosity. The true accretion
rate remains a factor of a few larger than this assumed rate throughout most
of the simulation. The rise in the accretion rate visible at $t \sim 100 \:
{\rm yr}$ is a consequence of the same gravitational instability that leads
to the formation of the second sink particle. The local contraction of the spiral arm 
causes material to be pulled back in its orbit, towards the collapsing region, where 
it collides with other material. This process causes a sufficient loss of angular 
momentum that the orbit can no longer be maintained and the gas moves inwards, 
resulting in an increased accretion rate onto the central protostar. 
{\it Right panel:} Evolution with time of the sink-mass
and accretion rate in  the simulation in which we assume an 
accretion rate of $10^{-2} \: {\rm M_{\odot} \: yr^{-1}}$. In this simulation,
the true accretion rate onto the sink remains smaller than the assumed value
throughout most of the simulation. It exceeds the assumed value only at very
early times after the formation of the first sink, although we caution that
the numerical resolution of this initial burst of accretion is poor, and that
we may be overestimating the rate at which it occurs. In this simulation, we 
also see a rise in the accretion rate associated with the formation of the second 
sink particle, but the magnitude of the increase is much smaller. This is a consequence
of the fact that the spiral structure close to the protostar is much less pronounced, owing
to the greater amount of heating produced by the increased accretion luminosity.
\label{accr_rate_2}}
\end{figure}

To determine $L_{\rm acc}$, and hence $\Gamma_{*}$, we need to
know the accretion rate onto the protostar, $\dot{M}_{*}$. Determining this 
self-consistently during the simulation is difficult, owing to the 
particle-based nature of our hydrodynamical model. The smallest amount of gas 
that can be accreted during any given hydrodynamical time-step is that corresponding 
to a single SPH particle, i.e.\ $10^{-5} \: {\rm M_{\odot}}$, and hence accretion onto
the sink particle proceeds as a series of discrete increments, rather than
a smooth increase in the sink particle mass. To obtain an accurate value for
the accretion rate, we therefore need to average over a large enough number
of time-steps to wash out this artificial discreteness. In the simulation discussed
in the main paper, we achieve this by recording the mass accreted by the sink 
particle during each hydrodynamical time-step, and determine the current value of $\dot{M}_{*}$
by computing a smoothed average of the accretion rate over the previous 
ten years. For times less than ten years after the formation of the protostar, we average
over a shorter period, and take the instantaneous accretion rate at the time that the
sink particle forms to be $\dot{M}_{*} =  5 \times 10^{-2} \: {\rm M_{\odot} \: yr^{-1}}$,
an overestimate of the true value (see Figure~S\ref{accr_rate}). As the protostellar accretion rate typically
decreases over time, this procedure gives us a slight overestimate of the true rate, as
can be seen in Figure~S\ref{accr_rate}, where we compare our estimated accretion rate (denoted 
by the black line) with the actual accretion rate onto the protostar (magenta line), which we measure
during a post-processing step.
In order to explore the sensitivity of our results to our method for determining $\dot{M}_{*}$,
we have also performed two simulations in which the accretion rate used within our accretion
luminosity calculation is not determined self-consistently during 
the simulation, but is instead maintained at a fixed value throughout. In these two simulations,
we set $\dot{M}_{*} = 10^{-3} \: {\rm M_{\odot} \: yr^{-1}}$ and $\dot{M}_{*} = 10^{-2} \: {\rm 
M_{\odot} \: yr^{-1}}$, respectively, as these values bracket the true value of  $\dot{M}_{*}$
during almost all of the period of time simulated, as illustrated in Figure S\ref{accr_rate_2}.
These simulations produce qualitatively and quantitatively similar results to our primary 
simulation, as demonstrated in Section~\ref{acc_rate_comp} below.

Finally, we note that
in our current simulations, we do not account for the photodissociation of H$_{2}$
molecules in the surrounding gas by radiation from the central protostar. We justify
this omission by noting that during the adiabatic accretion phase, the protostar has
a very large radius (see Eq.~\ref{eq:rstar} above), and hence a low effective temperature.
Therefore, only a very small fraction of the bolometric accretion luminosity of the 
protostar is emitted in the form of ultraviolet photons energetic enough photodissociate
H$_{2}$. As an illustrative example, consider the case of an $0.5 \: {\rm M_{\odot}}$
protostar accreting at a rate $\dot{M}_{*} = 10^{-3} \: {\rm M_{\odot} \: yr^{-1}}$: from
Eqs.~\ref{eq:rstar}--\ref{eq:lacc2} above, we find that the radius and luminosity of the 
protostar would be roughly $20 \: {\rm R_{\odot}}$ and $720 \: {\rm L_{\odot}}$,
respectively, from which it follows that the effective temperature of the
protostar at this point is roughly $6500 \: {\rm K}$ (where we have approximated the
protostellar emission as being from a pure black-body). The fraction of the protostellar
luminosity that falls within the Lyman-Werner bands of H$_{2}$ is therefore 
roughly $10^{-8}$, and so the accreting protostar emits roughly $10^{39} \:
{\rm photons} \: {\rm s^{-1}}$ within the Lyman-Werner bands. Even if we assume that 
every single one of these photons is absorbed and causes the dissociation of an 
H$_{2}$ molecule, the rate at which H$_{2}$ is destroyed by photodissociation is
less than $10^{-10} \: {\rm M_{\odot}} \: {\rm yr^{-1}}$, and hence only a negligible 
amount of H$_{2}$ would be destroyed during the period covered by our simulations.

\section{Initial conditions}
We begin our study with a cosmological simulation. We model the evolution of dark matter in a box with a side length of 200 comoving kpc. The cosmological simulation is initialized at $z=99$ with a standard $\Lambda$CDM fluctuation power spectrum, with a matter density $\Omega_{\rm m} =  1 - \Omega_{\Lambda} = 0.3$,  baryon density $\Omega_{\rm b} = 0.04$, Hubble parameter $h = H_{0} / 100 \: {\rm km \: s^{-1} \: Mpc^{-1}} = 0.7$, spectral index $n_{s} = 1.0$ and normalization $\sigma_{8} = 0.9$ \cite{spergel03}. Following the zoom technique used in Greif et al.\ \cite{gr10}, we first run a coarse dark matter (DM) simulation with $128^3$ particles and a particle mass of $\simeq 150~{\rm M}_\odot$ until the first minihalo exceeding a mass of $5\times 10^5~{\rm M}_\odot$ collapses. We then reinitialize the simulation at $z=99$ with additional small-scale power in a cube with $50$~kpc (comoving) on a side, centered on the location of the minihalo formed in the previous step, and
  replace each DM particle in the original simulation with $512$ DM and  512 gas particles. Outside of this region, a sufficiently large buffer zone is created, where the particle mass increases in factors of eight until the mass resolution of the original DM simulation is obtained. The final DM and gas particle masses in the high resolution region are $\simeq 0.3$ and $0.04~{\rm M}_\odot$, respectively. We then re-run the cosmological simulation, evolving the gas and dark matter until the gas in the minihalo has reached a number density of $10^{6} \: {\rm cm^{-3}}$, by which point the gas has gravitationally decoupled from its parent minihalo and has begun to collapse in its own right. This occurs at
a redshift $z \sim 17$, at which time the virial mass of the dark matter halo is $3.3 \times 10^{5} \: {\rm M_{\odot}}$. Note that this value is smaller than the mass quoted above because the runaway collapse of the gas cloud occurs slightly earlier than the time at which the minihalo was identified in the DM-only simulation.

At this point we discard the full cosmological simulation and focus our calculation on the central collapsing region and its immediate surroundings. These then become the initial conditions for the second phase of our study. The initial gravitational instability that leads to the collapse in the baryons occurs at a number density of around $n \sim 10^{4}$ cm$^{-3}$, in gas that has a mean temperature $T \sim 270$ K, implying that the Jeans mass at this point is around 350 M$_{\odot}$. At the point where we discard the cosmological simulation, the central density is around $10^{6}$ cm$^{-3}$, and so the collapse has already evolved over two orders of magnitude in density from the point at which the original instability occurred. To ensure that we capture the entire collapsing fragment in our simulations, and to avoid any unphysical boundary effects, we select a spherical region containing 1000~M$_{\odot}$ of gas to be re-simulated at higher resolution, and discard the rest of the SPH particles. We also discard the dark matter particles from the simulation at this point, as the further collapse of the gas is dominated by the self-gravity of the gas, rather than the gravitational potential of the dark matter, since it contributes only $\sim$ 70 ${\rm M}_\odot$ in the re-refined region. As such, removing the dark matter only changes the ratio of rotational to gravitational energy by around 7 percent in the region of interest, which is negligible given the variation expected from cosmic variance.

To account for the effects of  the missing gas  that should surround this central core, we include a external pressure term that modifies the standard gas-pressure contribution to the Gadget 2 momentum equation,
\begin{equation}
\frac{d v_{i}}{d t} = - \sum_{j} m_{j} \left[
f_{i}\frac{P_{i}}{\rho_{i}^{2}} \nabla_{i} W_{ij}(h_{i})
+ f_{j}\frac{P_{j}}{\rho_{j}^{2}}\nabla_{i} W_{ij}(h_{j}) \right] ,
\end{equation}
by replacing $P_{i}$ and $P_{j}$ with $P_{i} -  P_{\rm ext}$ and $P_{j} -  P_{\rm ext}$  respectively, where $P_{\rm ext}$ is the external pressure, and all other quantities have the usual meaning \cite{Springel05} . The pair-wise nature of the force summation over the SPH neighbors ensures that $P_{\rm ext}$ cancels for particles that are surrounded by other particles. However at the edge of the cloud, the extra pressure term does not disappear,  and thus mimics the pressure contribution from a surrounding medium. We use the average density and temperature at the edge of our cloud to define the value of  $P_{\rm ext}$.

In order to allow us to follow the collapse of the gas up to very high densities, it is necessary to substantially increase the resolution. The SPH particle mass in the original cosmological simulation was $0.04 \: {\rm M_{\odot}}$, and so the re-simulated region contains $25,000$ SPH particles. To increase the resolution, we `split' each of these particles into 100 new SPH particles of lower mass, giving a new particle mass at the beginning of our re-zoomed simulation of $4 \times 10^{-4}$ M$_{\odot}$. 

The particle splitting is done by randomly placing the sibling particles inside the smoothing length of the parent particle \cite{kitsionas2002, BL2003, yoh08}. Apart from the mass of the siblings, which is 100 times less than that of the parent, they inherent the same values for the entropy, velocities and chemical abundances as their parents, thereby ensuring conservation of mass, linear momentum and energy during the splitting of the particles. Contrary to claims in the literature, splitting the particles does not conserve angular momentum by construction, since the mass distribution changes during the split (and adaptive mess refinement codes suffer from the same problem). However we find the error in the angular momentum introduced during the splitting to be small, around a percent.   
This process is repeated several times during the simulation to ensure that the resolution in the collapsing region remains high. However, in these subsequent `splits', the particle mass is only changed by a factor of five at most. In Table~S\ref{tab:split}, we list the densities at which we split the
particles, the temperature and local Jeans mass of the gas at the point where splitting occurs, the total mass of gas that undergoes particle splitting at each stage, and
the particle mass before and after the splitting. We also list the number of SPH particles per Jeans mass and per Jeans length prior to the splitting, allowing us to demonstrate that we always apply our splitting technique to highly resolved regions. 

Our particle splitting technique ensures that particles of significantly different mass (i.e.\ those with mass ratios greater than 5) never come in contact with one another during the final stages of the simulation, minimizing any potential problems due to the mixing of SPH particles with wildly different masses. The fact that the inner 82 M$_{\odot}$ of gas is resolved with a uniform particle mass of $10^{-5}$ M$_{\odot}$ also ensures that the disk is formed by SPH particles that are of equal mass, and we halt the simulations long before any of the higher-mass particles have reached the disk. 

At the point of disk fragmentation, our calculations have in excess of 300,000 SPH particles in the disk. Even for mildly self-gravitating disks, this has been shown to be sufficient to provide converged evolution for the angular momentum transport \cite{lr2004, lr2005}. Disks that are more strongly self-gravitating, such as the one we present in this Report, are easier to resolve, as they have a larger scale height.

It should also be stressed that rather than promoting artificial fragmentation, the `kernel' softening employed within Gadget-2 to soften gravitational forces on small scales will tend to suppress fragmentation in unresolved regions rather than promote it \cite{bb97, Whitworth1998, Hubber2006}, since the gravitational forces will be diluted to scales longer than the pressure forces. The high resolution that we employ in this study ensures that we capture all of the fragmentation in the density regime in which we have a reliable thermodynamic model for the gas, while in the density regime in which our thermodynamical model begins to break down (i.e. $n > 10^{17}$ cm$^{-3}$), the gas is stable to further fragmentation. 

\begin{table}
\caption{We summarize here the details of SPH particle masses that result from the `particle splitting' that is used to increase the spatial resolution in our simulation. The quantities given are: the number density at which the resolution is increased; the mean temperature at this density; the corresponding Jeans mass; the mass in the region that is refined; the particle mass, before and after the split; the number of particles resolving a Jeans mass; and the number of particles per Jeans length.  
%Note that the Jeans mass ($m_{\rm J}$) and Jeans length $\lambda_{\rm J}$ used here are those for a uniform sphere, and so the number of particle in a cube of side $\lambda_{\rm J}$ is $6/\pi$ larger than the value quoted here. 
For comparison, the bottom two rows show the densities and temperatures associated with the onset of gravitational instability in the disk and the formation of sink particles.
\label{tab:split}}
\begin{center}
\begin{tabular}{c|c|c|c|c|r|r}
n		     &  T    & $m_{\rm J}$ & $M_{\rm split}$ & $M_{\rm part}$ &	$N_{\rm part}/M_{\rm J}$  & 
$N_{\rm part}/\lambda_{\rm J}$ \\
(cm$^{-3}$) &  (K) & (M$_{\odot}$)     &  (M$_{\odot}$)         & (M$_{\odot}$)          & \\
\hline
\hline
$10^{6}$   & 350 & 52   &           & 0.04 & 1,304	     &  11        \\
$10^{6}$      & 350    & 52    & 1000 & $4 \times 10^{-4}$ & 130,400 &  50       \\
\hline
$10^{10}$ & 750 & 1.6  &		  & $4 \times 10^{-4}$ &	4,090     &  16        \\
$10^{10}$ & 750 & 1.6  &  336	  & $1 \times 10^{-4}$ &	16,360   &  25       \\
\hline
$2 \times 10^{10}$     & 780    &  1.2     &          & $1 \times 10^{-4}$ &  12,269   &  23        \\
$2 \times 10^{10}$     & 780    & 1.2      & 178  & $5 \times 10^{-5}$&  24,539   &  29        \\
\hline
$4 \times 10^{10}$    & 790 & 0.880 &          &  $5 \times 10^{-5}$&  17,687   &  26       \\
$4 \times 10^{10}$	   & 790 & 0.880 & 82    &  $1 \times 10^{-5}$ &  88,433   & 44     \\
\hline
\hline
$10^{14}$ &1200& 0.033 &   	   & $1 \times 10^{-5}$ &  3,311     & 15        \\
$10^{17}$ &1400&0.001	 &         & $1 \times 10^{-5}$ &  132      &    5         \\
\end{tabular}
\end{center}
\end{table}

\section{State of the cloud at the onset of star formation}
One potential concern about our findings is that since we have examined only 
a single realization of primordial protostellar collapse, we may be obtaining 
results that are not typical of Population III star formation in general. If 
the minihalo that we have selected to examine is unusual in some way, then 
the results that we have obtained from it may be misleading. Ultimately, this
is a point that will need to be addressed by performing a large set of 
simulations along the lines that we have laid out here, allowing many 
different realizations to be examined, but at present we do not have 
sufficient computational resources for such a large-scale study. 
However, as we argue in the main article, we can gain confidence that our
simulated halo is not atypical by comparing its properties to those 
reported in other studies of Population III star formation. 

As mentioned above, the initial gas distribution for our high resolution
SPH simulation was taken from a minihalo with a  dark matter virial mass of 
$3.3 \times 10^{5} \: {\rm M_{\odot}}$ and a dimensionless spin parameter 
$\lambda = 0.034$. Runaway gravitational collapse of gas within the minihalo 
occurred at a redshift $z \sim 17$. For comparison, the halos investigated in two
of the most recent high-resolution studies of Population III star formation 
\cite{yoh08,tao09} had masses of $5 \times 10^{5} \: {\rm M_{\odot}}$ and
$5.8 \times 10^{5} \: {\rm M_{\odot}}$, respectively, and collapsed at redshifts 
$z=14$ and $z=19$, respectively. Our halo mass is therefore slightly smaller
than these previous values, while our collapse redshift is well within the range
found in previous work. Our halo spin parameter is slightly smaller than the
value of $\lambda = 0.042$ quoted in Turk et~al.\ \cite{tao09}, but lies very close to the
mean of the spin parameter distribution found in previous studies of high-redshift
dark matter halos \cite{jch01,dn10}.

\begin{figure}
\includegraphics[scale=0.8]{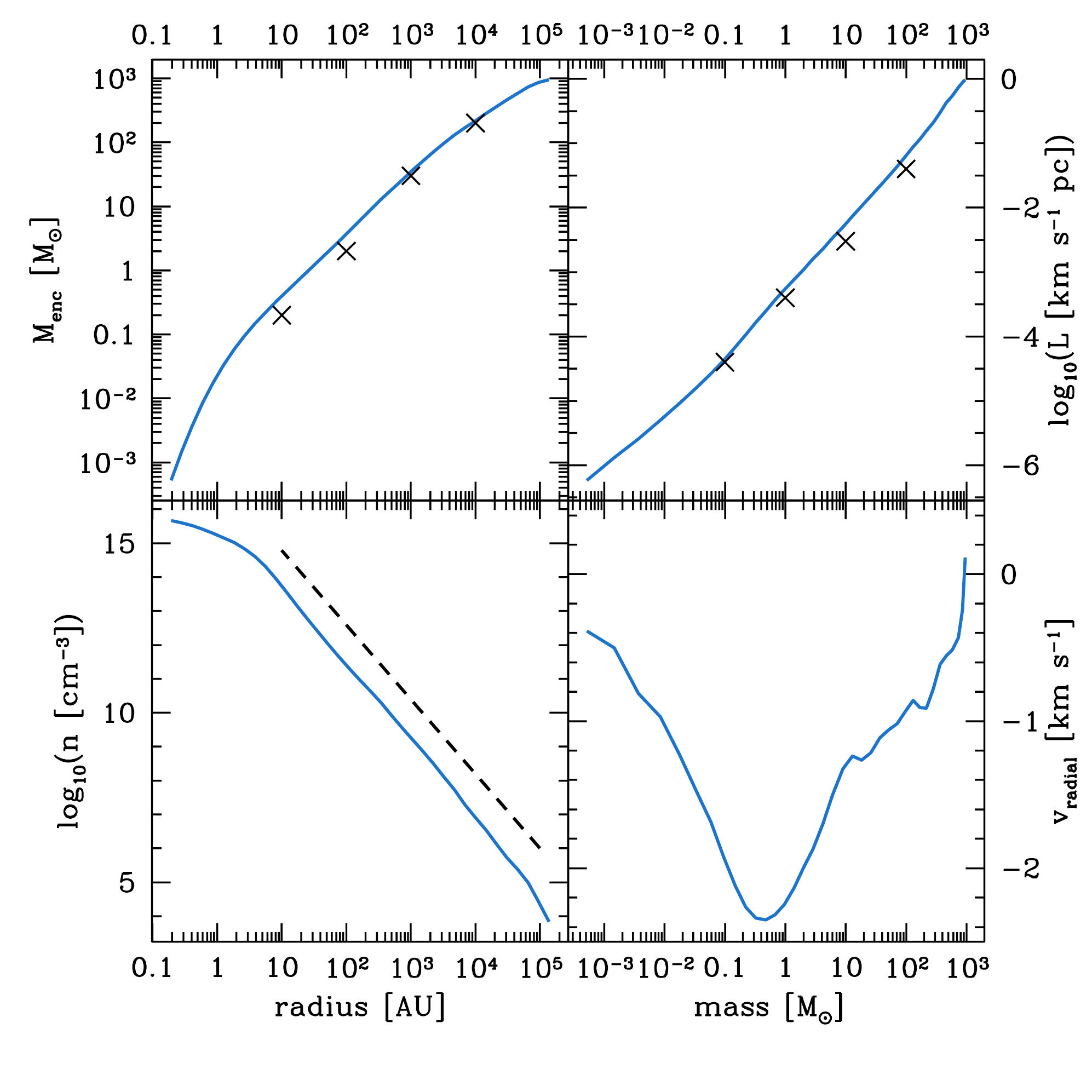}
\caption{{\it Upper left panel:} mass enclosed as a function of the radial distance
from the densest point in the simulation, plotted just before the formation of 
the first sink particle. The black crosses represent the corresponding values in the
simulation presented by Abel et~al.~\cite{abn02}. {\it Upper right panel:} spherically averaged value of the
specific angular momentum, $L$, plotted at the same time. Crosses represent the values
of $L$ for the indicated enclosed masses in the simulation presented by Abel et~al.~\cite{abn02}.
{\it Lower left panel:} spherically averaged number density profile at the same time, 
plotted as a function of the enclosed mass.  To help guide the eye, the dashed line represents the  
$\rho \propto r^{-2.2}$ slope found in previous simulations of Population III star formation.
{\it Lower right panel:} spherically averaged radial velocity profile
\label{presink}}
\end{figure}

We can also compare the state of the gas immediately prior to the formation
of the first sink particle with previously reported results. In Figure S\ref{presink}, 
we show plots of the enclosed mass $M_{\rm enc}$ and specific angular momentum 
$L$ of the gas as a function of distance from the centre of  the collapse (left-hand panels), 
and of the number density $n$ and radial velocity $v_{\rm radial}$ of the gas as a function 
of the enclosed gas mass (right-hand panels). If we compare these radial profiles to those 
reported in previous high-resolution studies \cite{abn02,yoha06,yoh08,tao09}, we find 
reasonable agreement. We recover the standard $\rho \propto r^{-2.2}$ density profile at radii 
greater than a few AU, and masses above $0.1 \: M_{\odot}$. 
Our values for the specific angular momentum agree very well with the values reported elsewhere 
\cite{abn02,yoha06}, demonstrating that our gas is not rotating artificially rapidly.
We also recover a similar infall velocity profile as in previous studies, with a peak value of 
about $2.5 \: {\rm km} \: {\rm s^{-1}}$, again in good agreement with previously reported results
\cite{abn02,yoh08}. We do not have sufficient resolution to recover the post-shock, 
sub-AU scale infall visible in Fig.~3B of Yoshida et~al.\ \cite{yoh08}: this would occur within the region 
that is represented by our central sink  particle.

\begin{figure}
\includegraphics[scale=0.7]{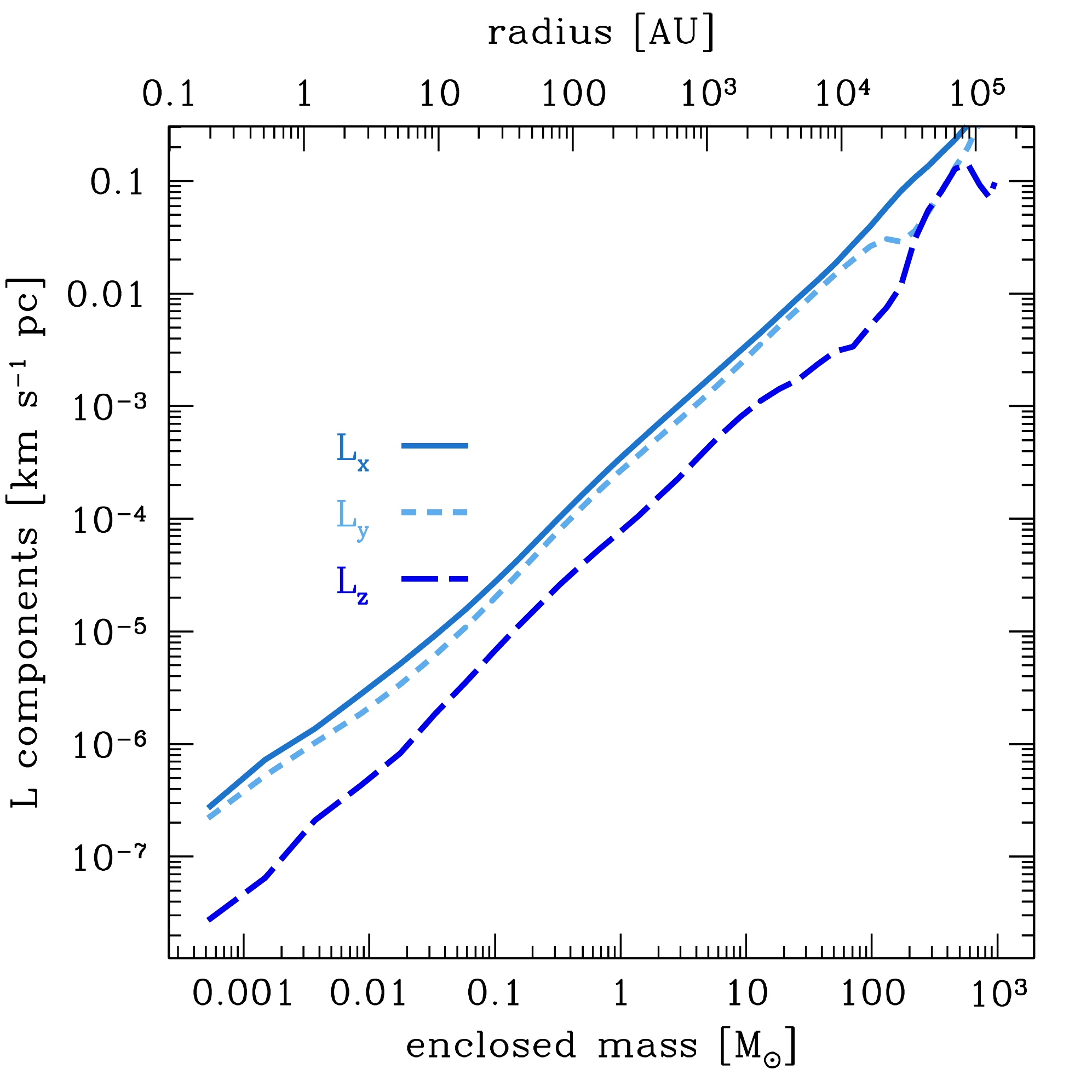}
\caption{The magnitudes of the individual components of the specific angular momentum vector
in each shell, plotted as a function of both enclosed mass and radius. The values are mass-weighted 
and are taken from a point in the calculation immediately prior to the formation of the first sink particle.
The angular momentum is calculated with respect to the densest point in the gas.
\label{angmom}}
\end{figure}

In Figure~S\ref{angmom}, we show the radial profile of the three components of the angular 
momentum vector at a point in the cloud's evolution just before the creation of the first
sink particle. The fact that the relative magnitudes of each component remain constant 
over such a large distance (and enclosed mass) in the cloud shows that the mass shells are, 
for the most part, rotating around a common axis. Only as the enclosed mass approaches a 
value roughly comparable to the original Jeans mass in the minihalo, $M_{\rm J, init} = 350 \:
{\rm M_{\odot}}$, do we see a significant
change in the orientation of the angular momentum vector. This shows that the collapsing region 
had a well defined rotation axis,  while the large-scale motions in the minihalo were much more 
chaotic. It would be interesting to compare these results to those from previous studies, but to
the best of our knowledge, the required data is not available in the astrophysical literature. 

\begin{figure}
\includegraphics[scale=0.7]{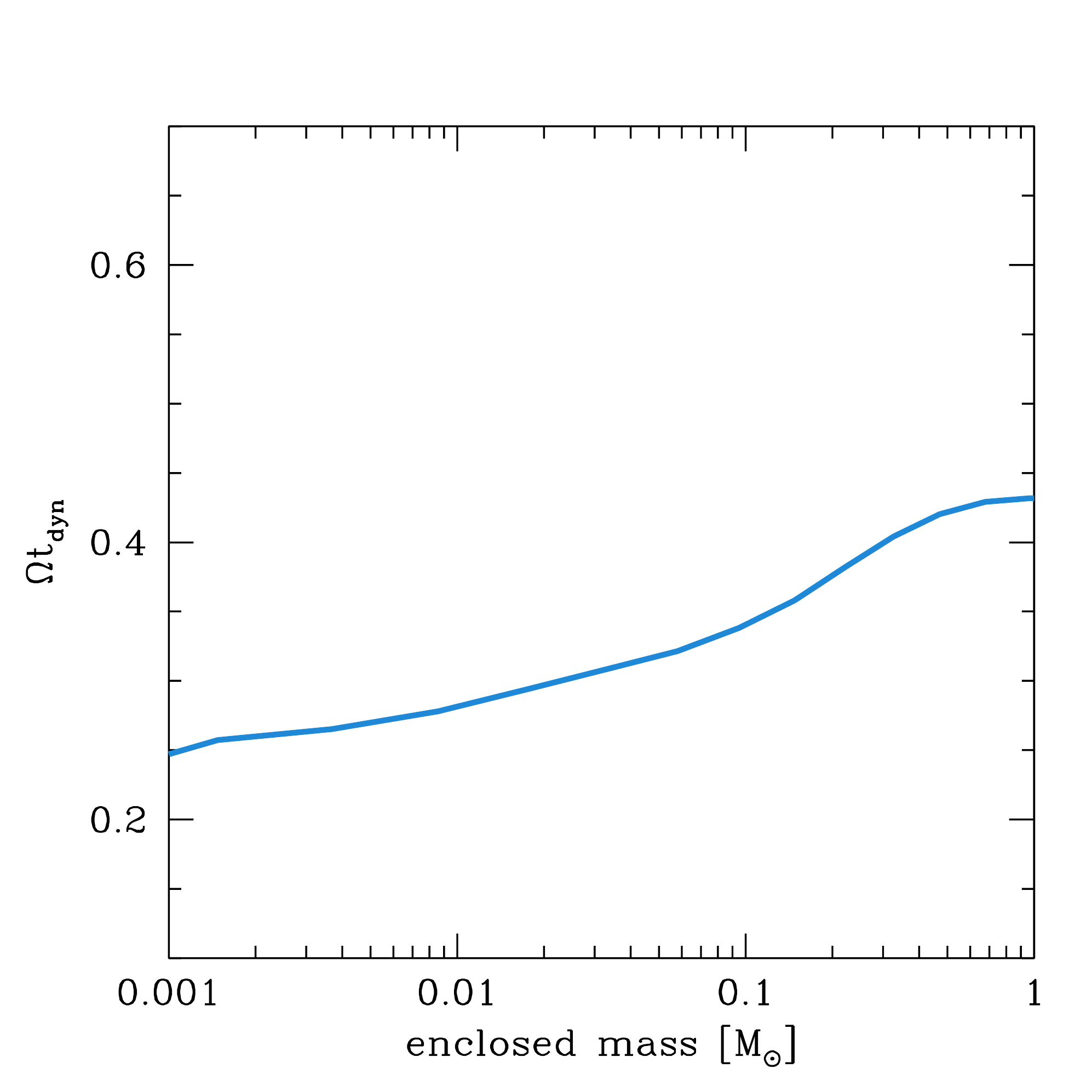}
\caption{The product of the mean angular velocity $\Omega$ and the local dynamical time $t_{\rm dyn} = 
1 / \sqrt{4 \pi G \rho}$, plotted as a function of the enclosed gas mass, at a point in the calculation 
immediately prior to the formation of the first sink particle.
 \label{omega_t}}
\end{figure}

We have also examined how the product of the mean angular velocity $\Omega$ and the local dynamical 
time $t_{\rm dyn} = 1 / \sqrt{4 \pi G \rho}$ evolves as we move away from the site where the first protostar
forms (Figure~S\ref{omega_t}). Yoshida et~al.\ \cite{yoh08} measured this for the central $0.01\: {\rm M_{\odot}}$
of gas in their simulation at the point at which they could no longer follow the evolution of the system, and 
found a value $\Omega t_{\rm dyn} = 0.25$, in good agreement with our own results. 

\begin{figure}
\includegraphics[scale=0.7]{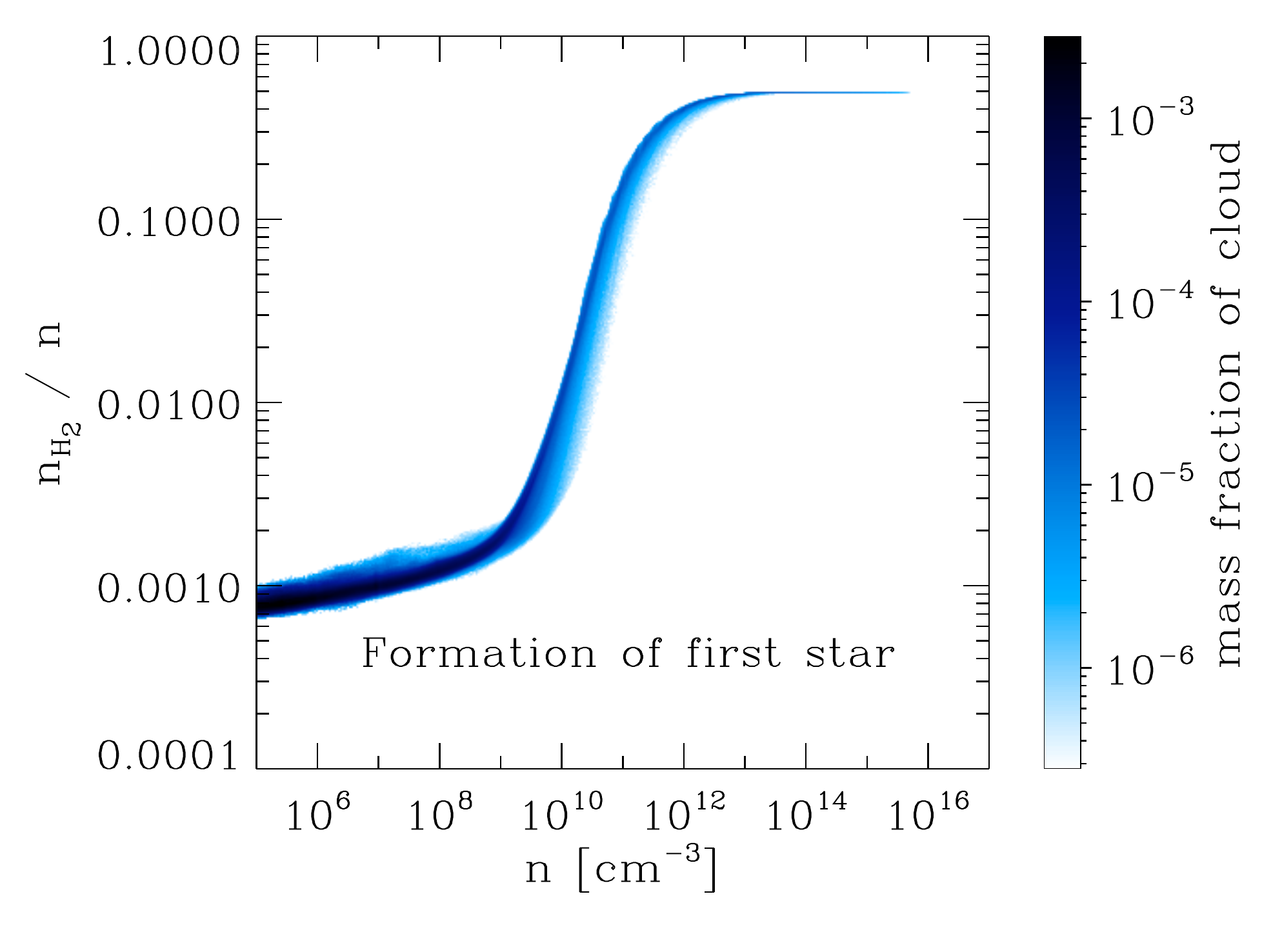}
\caption{\label{fig:h2fraction}  
Ratio of the H$_{2}$ number density, $n_{\rm H_{2}}$, to the number density of hydrogen nuclei, $n$, 
plotted as a function of $n$, at a time immediately prior to the formation of the first sink particle. Note 
that a value $n_{\rm H_{2}} / n = 0.5$ indicates fully molecular gas.}
\end{figure}

Finally, we have also examined the H$_{2}$ distribution in the cloud just before the formation of the first sink
particle. In Figure~\ref{fig:h2fraction}, we show how the H$_{2}$ fraction (defined here as the ratio of
the H$_{2}$ number density $n_{\rm H_{2}}$ to the number density of hydrogen nuclei, $n$) varies as
a function of $n$. We see that the transition from atomic to molecular hydrogen due to three-body H$_{2}$
formation occurs at a density of roughly $10^{10} \: {\rm cm^{-3}}$, in line with the finding of previous
studies that have used the same three-body H$_{2}$ formation rate coefficient \cite{abn02,tao09}.

\section{Sensitivity to the choice of accretion rate}
\label{acc_rate_comp}
Once gas begins to accumulate in an accretion disk around the initial sink particle, its
subsequent evolution will depend to some extent upon its temperature, and hence on 
the degree to which it is heated by radiation from the accreting protostar. As already 
discussed in Section~\ref{method}, one of the difficulties involved in a self-consistent
determination of the accretion luminosity of the protostar lies in establishing an 
appropriately time-averaged estimate for the protostellar accretion rate that is insensitive
to the particle-based nature of SPH. In our present study, we calculate the instantaneous 
accretion rate onto each protostar by computing a smoothed average of the accretion rate
over the previous ten years. This procedure yields a reasonable estimate of the true
accretion rate, as Figure~S\ref{accr_rate} demonstrates, but nevertheless it remains an approximation.
It is therefore informative to examine the sensitivity of our results to the value of the 
protostellar accretion rate, in order to help us understand whether the approximation 
that we are making is likely to be a significant source of uncertainty. 

\begin{figure}
\includegraphics[scale=0.7]{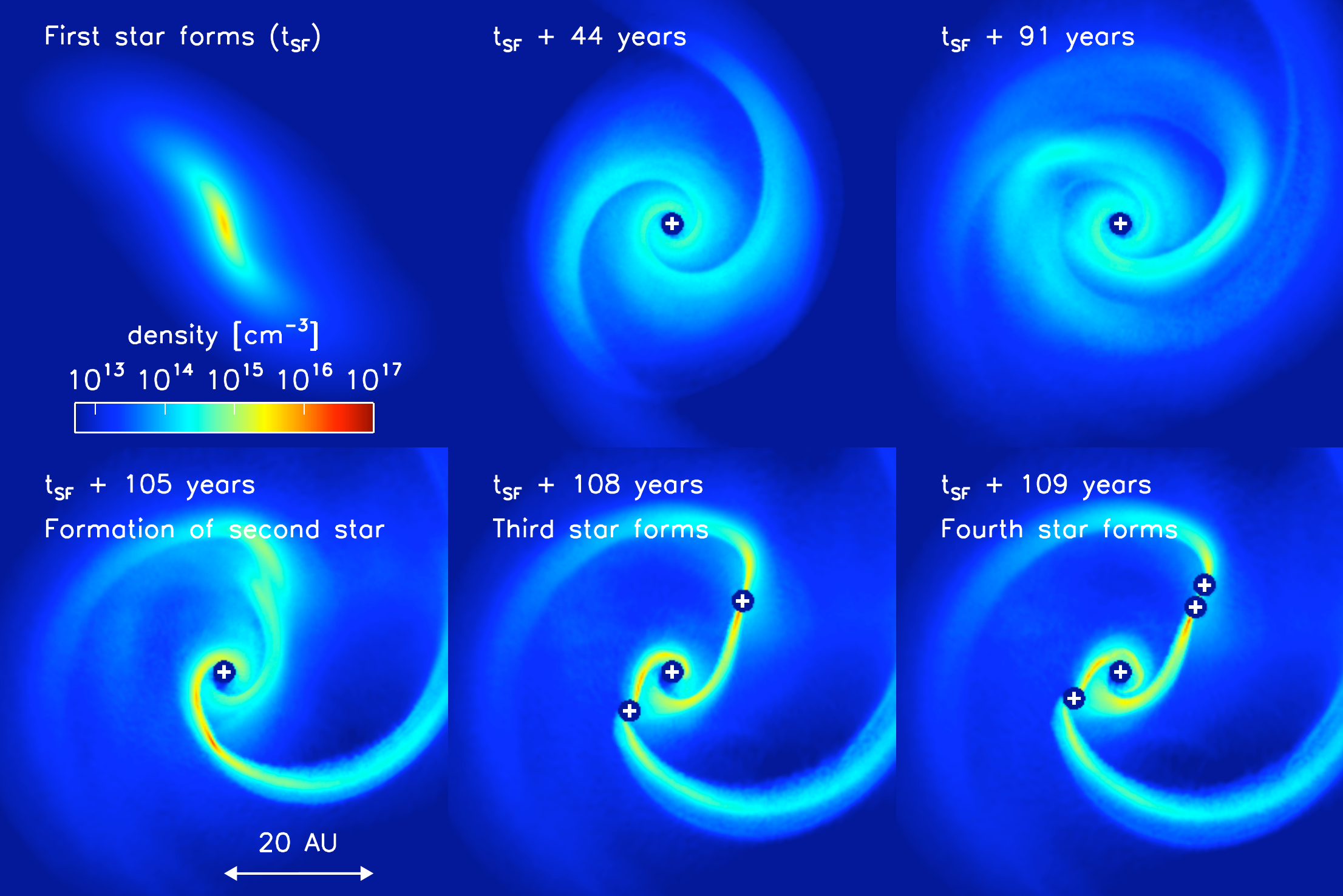}
\caption{The evolution of the density in the $\dot{M}_{*}  = 10^{-3} \: {\rm M_{\odot} \: 
yr^{-1}}$ simulation, showing the build-up of the accretion disk around the central 
protostar. Note that the scale and color table are different from those used in Figure 1 in the 
main article.
\label{cartoon_lowmdot}}
\end{figure}

\begin{figure}
\includegraphics[scale=0.7]{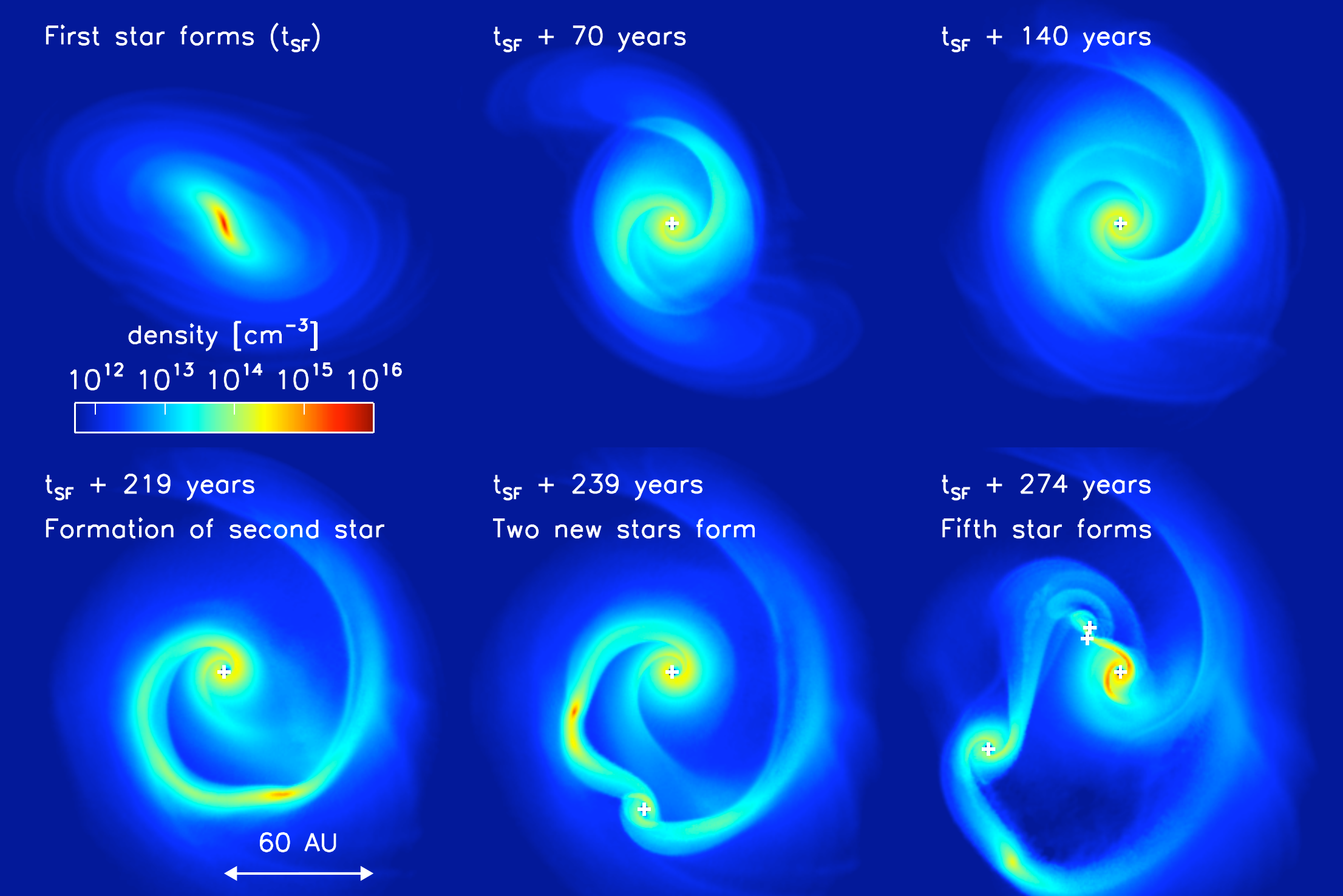}
\caption{The evolution of the density in the $\dot{M}_{*}  = 10^{-2} \: {\rm M_{\odot} \: 
yr^{-1}}$ simulation, showing the build-up of the accretion disk around the central 
protostar. Note that the scale is different to that used in Figure 1 in the main
article.
\label{cartoon_highmdot}}
\end{figure}

To address this issue, we have examined the results of two additional simulations, performed
using fixed values of $\dot{M}_{*} =10^{-2} \: {\rm M_{\odot} \: yr^{-1}}$ and 
$\dot{M}_{*} =10^{-3} \: {\rm M_{\odot} \: yr^{-1}}$, respectively, for the protostellar accretion 
rate used to compute the accretion luminosity of the protostar. As we can see from
Figure~S\ref{accr_rate}, these two values  bracket the true accretion rate during the 
period of time covered by our simulations. We find evidence for disk fragmentation in both
of these simulations. When the accretion rate onto the central protostar is low, and hence
the accretion luminosity is less effective, fragmentation occurs more rapidly than when the accretion rate 
is high. In the $\dot{M}_{*}  = 10^{-3} \: {\rm M_{\odot} \: yr^{-1}}$ simulation, it takes only 105~years before 
the disk fragments to form a second sink particle, whereas in the higher $\dot{M}_{*}$ simulation it takes
274~years for this to happen. Furthermore, when the accretion rate is low, fragmentation occurs closer
to the central protostar than when the accretion rate is high, as can be seen by comparing 
Figures~S\ref{cartoon_lowmdot} and S\ref{cartoon_highmdot} (although note the difference in the linear 
scale of these two figures). Nevertheless, in both cases,  the general features of the fragmentation 
are the same:  the symmetry in the spiral pattern breaks and one arm becomes gravitationally 
unstable, with part collapsing to form a new protostar. The second arm follows quickly after it 
(no more than a few tens of years in both cases), with the result that 4 to 5 stars are formed 
within a few years of the onset of fragmentation in the disk.

If we compare the results of these simulations with the results of the simulation that used a more
self-consistent determination of $\dot{M}_{*}$, we see that the behaviour of the latter closely resembles 
that found in the $\dot{M}_{*}  = 10^{-3} \: {\rm M_{\odot} \: yr^{-1}}$ simulation, with fragmentation 
occurring relatively quickly at a distance of roughly 20~AU from the central protostar. This result is not
particularly surprising, given that the true accretion rate onto the central protostar is only slightly larger
than $10^{-3} \: {\rm M_{\odot} \: yr^{-1}}$ at the point at which the disk fragments (Figure~S\ref{accr_rate}), but 
the fact that we still find that fragmentation of the disk occurs even when the adopted accretion rate is
substantially larger than this argues that any small uncertainties in our determination of  $\dot{M}_{*}$ 
are unlikely to be significantly influencing our results.

\begin{figure}
\includegraphics[scale=0.7]{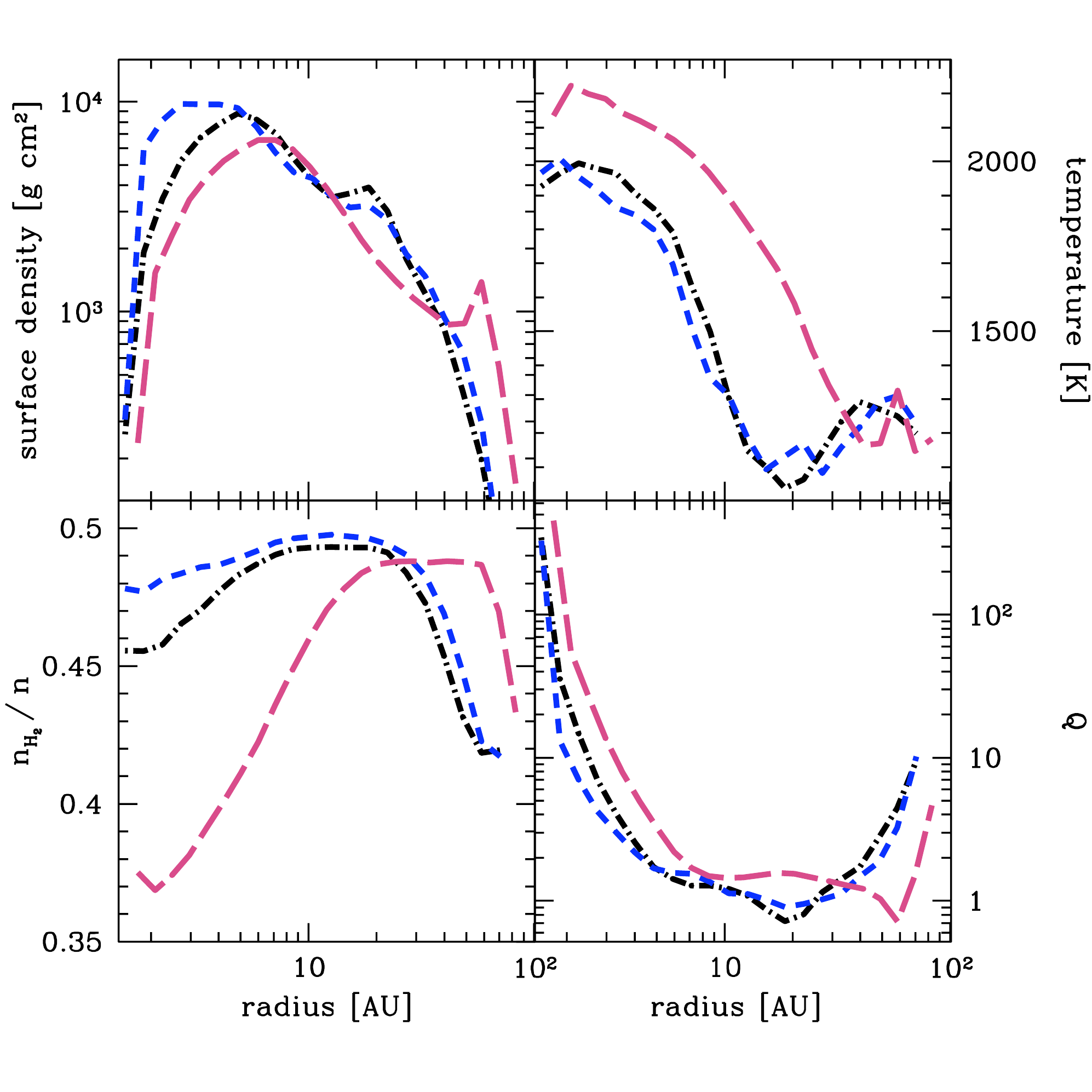}
\caption{Radially averaged disk properties, measured at a point immediately prior to the
formation of a second sink particle, in simulations with the live accretion rate (black, dot-dashed line),  $\dot{M}_{*} = 10^{-3} \: {\rm M_{\odot} \: yr^{-1}}$
(blue, short-dashed line) and $\dot{M}_{*} = 10^{-2} \: {\rm M_{\odot} \: yr^{-1}}$ (pink, long-dashed line). 
In the  low fixed $\dot{M}_{*}$ simulation, the second sink forms at $t_{\rm SF} + 105$~years, and in the
high fixed $\dot{M}_{*}$ simulation, the second sink forms at  $t_{\rm SF} + 274$~years, where $t_{\rm SF}$
denotes the time at which the initial sink particle formed. In the simulation with the live accretion rate, the second sink particle forms at $t_{\rm SF} + 91$~years, a little earlier than in the low fixed accretion rate calculation. 
\label{diskprops}}
\end{figure}

In Figure~S\ref{diskprops},  we compare the surface density, temperature, H$_{2}$ fraction 
and Toomre stability parameter for the disks in our three simulations at the point at which the 
second sink particle forms in each simulation. We see that when $\dot{M}_{*}$ is small, the accretion 
disk is systematically colder at all radii $R < 40$~AU, and is also significantly denser within the 
central 5~AU, although at larger radii the disk surface densities do not differ by a large factor.
The higher central temperature in the high $\dot{M}_{*}$ simulation causes a pronounced central 
dip in the H$_{2}$ fraction (defined here as $n_{\rm H_{2}} / n$, which means that a value of 0.5 
corresponds to fully molecular gas). This dip is almost absent in the other two simulations.
Comparing the Toomre parameter of the three disks, we see that they all have $Q \sim 1$ at radii 
greater than about 5--10~AU, indicating that in each case, the disk is strongly self-gravitating. It is 
therefore not surprising that we find fragmentation in all three simulations. In fact, the conditions in 
the disk and infalling envelope are similar to those found to lead to fragmentation in a recent study 
of present-day star formation \cite{krat10}.

%Finally, we note that the true accretion rates that we measure from the central sink particles
%during a post-processing step are roughly the same (Figure~S\ref{accr_rate}), demonstrating that the 
%system will always adjust to find a new equilibrium in the accretion rate, and that this equilibrium is closer 
%to a value of $10^{-3}$ M$_{\odot}$ yr$^{-1}$, than it is to $10^{-2}$ M$_{\odot}$ yr$^{-1}$. 

\section{Protostellar masses and accretion rates}
\begin{figure}
\centerline{
\includegraphics[scale=0.39]{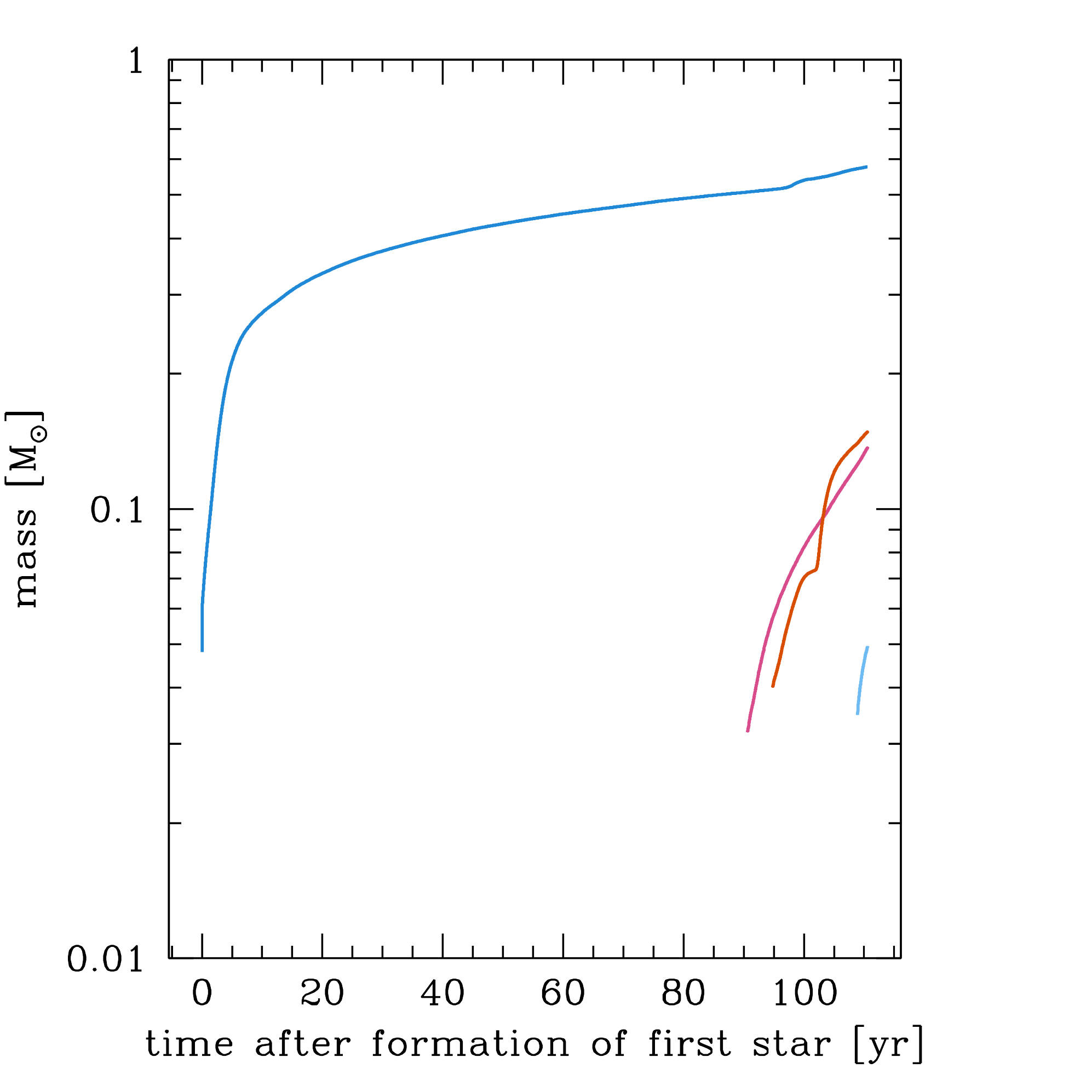}
\includegraphics[scale=0.39]{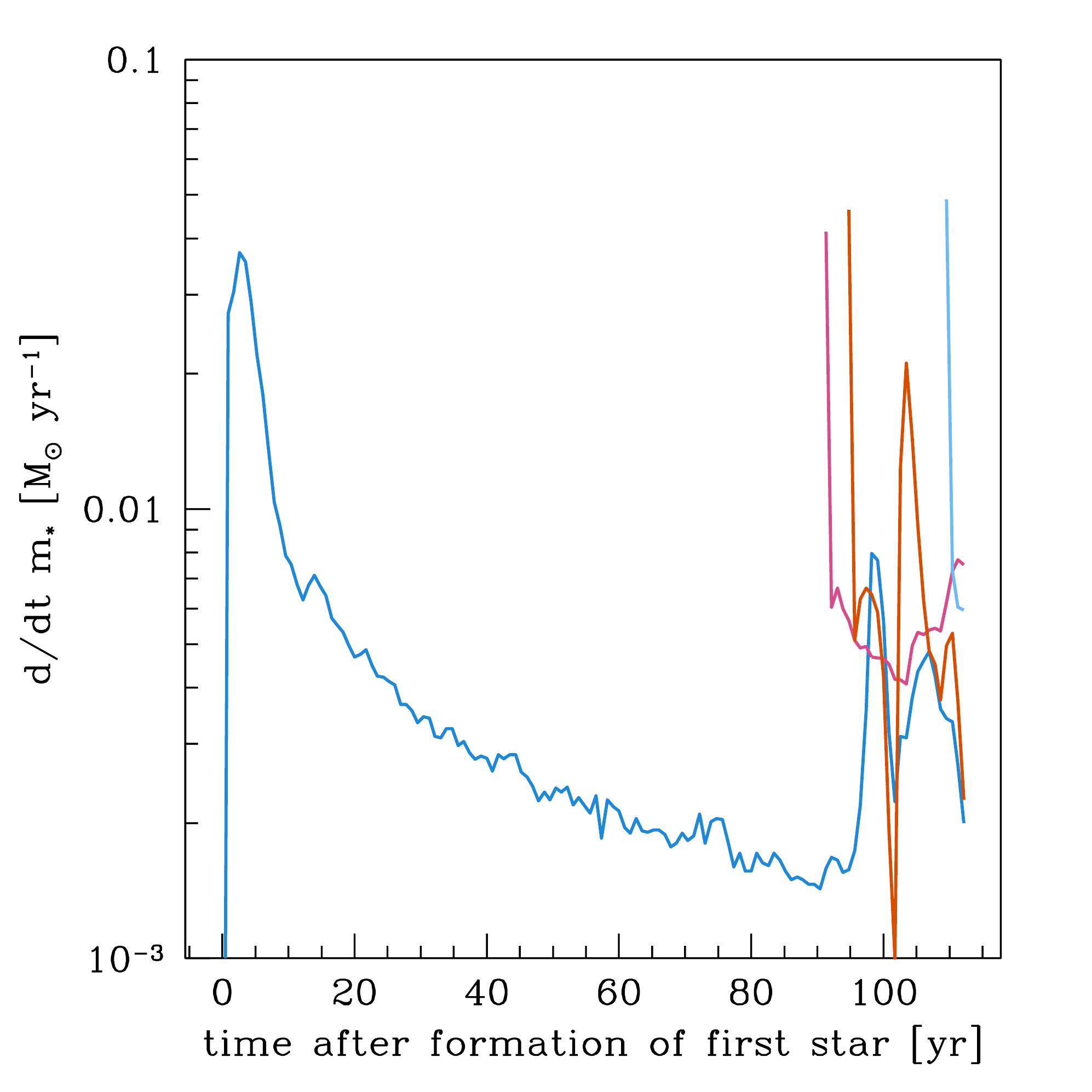}
}
\caption{\label{sinkmass} 
{\it Left panel:} The evolution with time of the masses of the sink particles.
{\it Right panel:} The accretion rates onto the individual sinks. Once the disk
begins to fragment, the strong gravitational torques allow more mass to flow
onto the central protostar. However, the complex interactions between the individual
sinks, and between the sinks and the disk cause the accretion rates of all of the
sinks to become highly variable.}
\end{figure}

In Figure~S\ref{sinkmass}, we show how the masses of the protostars evolve over the first 120 years 
of the simulation and how the accretion rates onto the protostars vary with time. The first point to note
is that all of the protostars have a mass of roughly $3 \times 10^{-2} \: {\rm M_{\odot}}$ when they form.
This is not particularly surprising, as this mass scale is simply set by the Jeans mass within the 
protostellar accretion disk. However, accretion from their surroundings rapidly increases the protostellar
masses, which typically exceed $0.1 \: {\rm M_{\odot}}$ within only 10--20 years of their formation.  
Another point to note is that although accretion onto the central protostar initially proceeds in a fairly 
smooth manner, the onset of fragmentation in the disk produces strong gravitational torques that allow
more mass to flow into the central protostar, but that also cause the accretion rate to become far more 
variable. Similarly, the complex gravitational interactions between the new protostars and the disk also
cause  the accretion rates onto these protostars to vary strongly with time. 

As the disk breaks up, the computational expense involved in following its further evolution becomes 
very large, and so for the present we have been forced to terminate our simulations once four or five
protostars have formed. At the point at which we stop the simulations, there is roughly an order of 
magnitude difference between the mass of the most massive and least massive protostars, but given
the high rates at which all of the protostars are accreting gas, plus the fact that at this point less than
0.1\% of the available gas has been accreted, we cannot state with any certainty what the final spread
in stellar masses will be, or indeed how many protostars will ultimately be formed. 

\section{Rotational support of the accretion disk}
\begin{figure}
\includegraphics[scale=0.7]{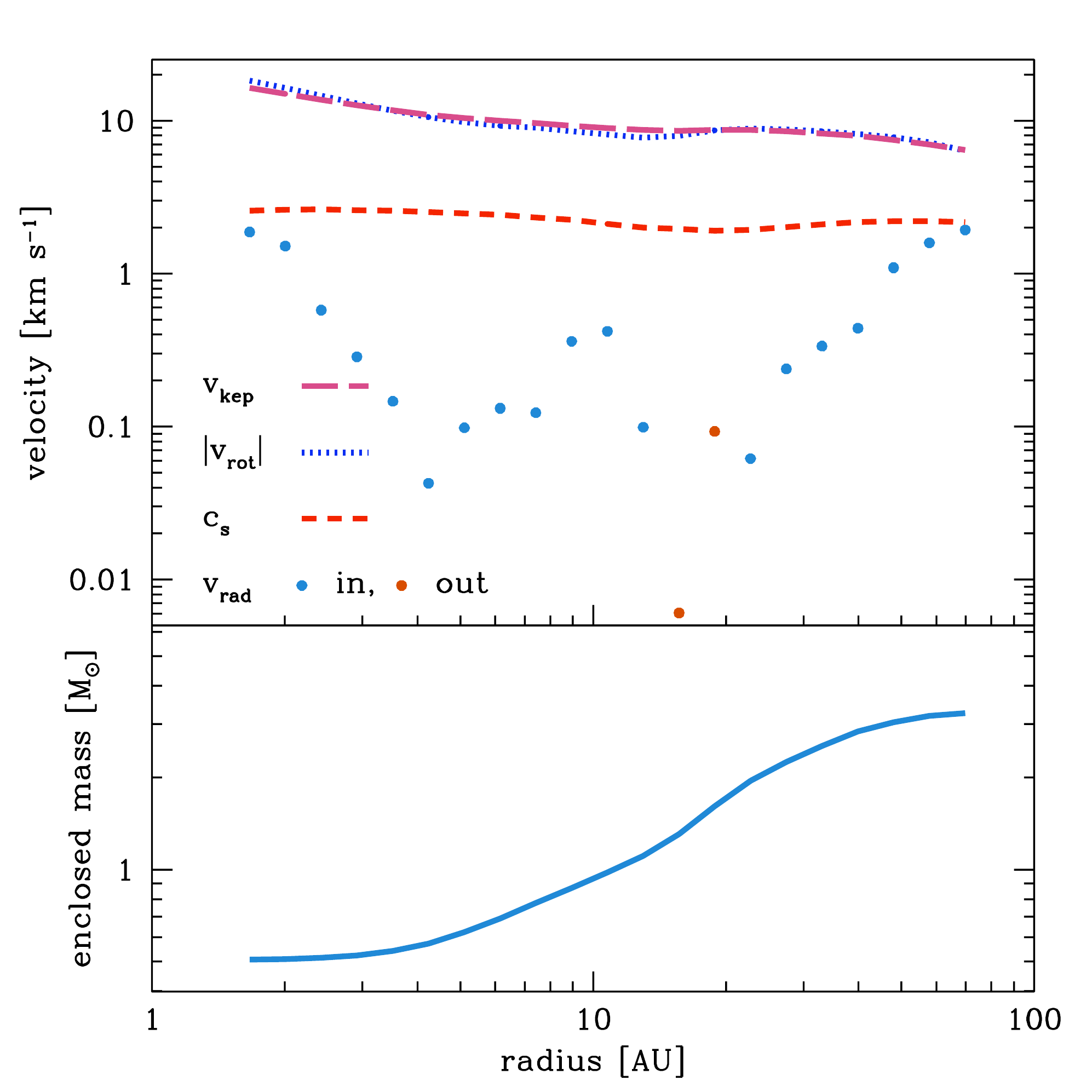}
\caption{{\em Upper panel:} radial velocity profiles in the plane of the disk, along with the 
sound speed of gas, at a point in the simulation just before the disk fragments to form 
a second protostar. Regions of the disk that are moving towards the central star are 
shown by blue dots, while those those moving away are shown in red. 
{\em Lower panel:} Mass profile of the protostar-disk system just before fragmentation.
\label{fig:rot}}
\end{figure}

The velocity and mass profiles in the protostellar disk at the point of fragmentation 
are shown in Fig S\ref{fig:rot}. We see that despite the strong spiral features, the disk is 
essentially Keplerian. We also see that a substantial fraction of the disk is moving {\em 
away} from the central protostar as the new fragment forms, demonstrating that the new 
protostar is forming in a region that is directly responsible for the transport of 
angular momentum through the disk. In fact, the fragmentation occurs at a radius of about 
20 AU from the central protostar: roughly the point at which the material moving outwards 
collides with the gas coming in from the envelope.

The mass profile of the protostar-disk system is also shown in Fig S\ref{fig:rot}. We see that 
the mass enclosed by the disk at 20 AU is roughly 2 M$_{\odot}$, while the mass in the 
central protostar at this time is only around 0.5 M$_{\odot}$. The disk is therefore 
significantly more massive than the central protostar, a feature commonly reported for the early phases of protostellar mass growth in the simulations of present-day star formation \cite{ban04}.

\section{Thermodynamics of the accretion disk}
\label{sec:thermo}

\begin{figure}
\includegraphics[scale=0.8]{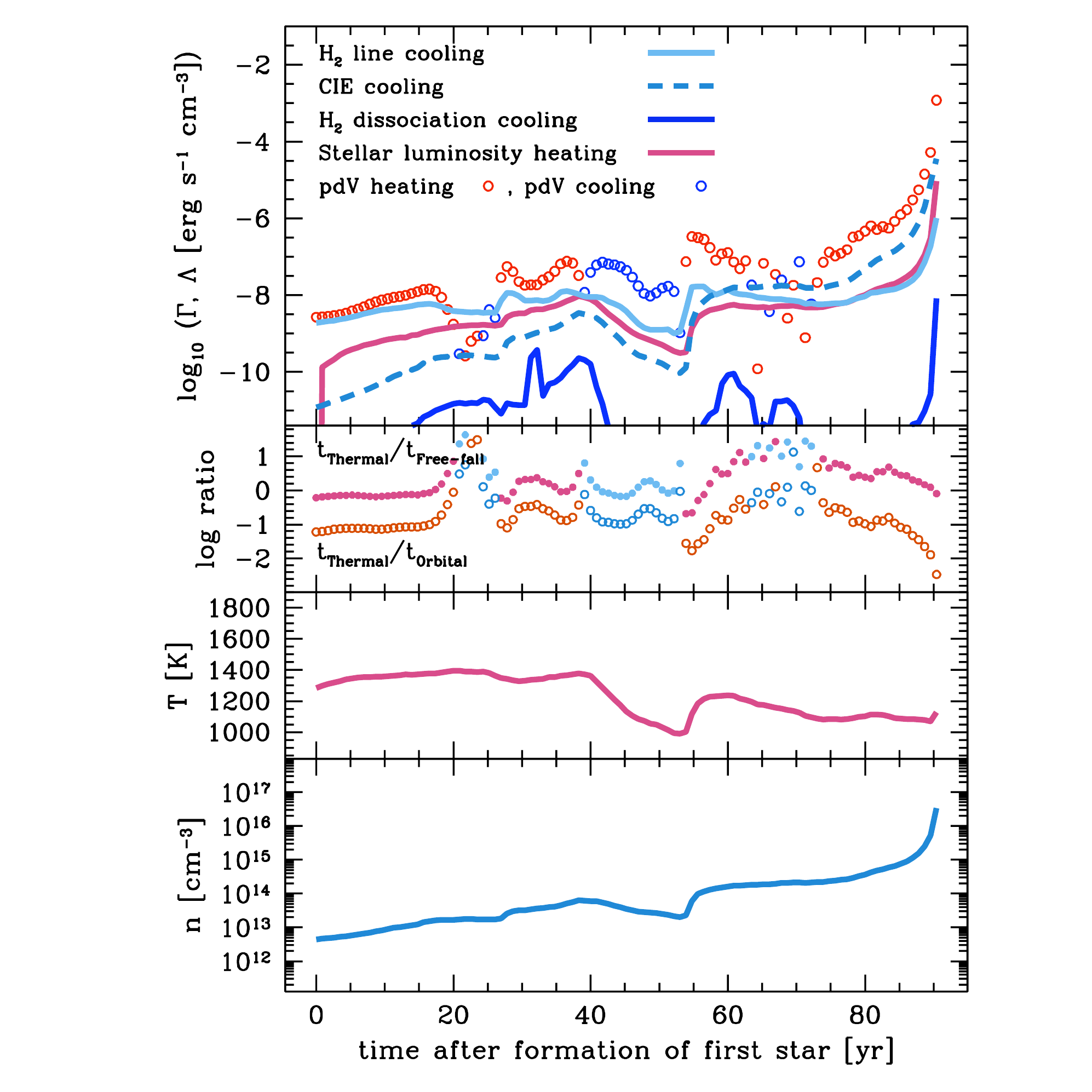}
\caption{(a) Dominant heating and cooling processes in the gas that forms the second sink particle.
(b) Upper line: ratio of the thermal timescale, $t_{\rm thermal}$, to the free-fall 
timescale, $t_{\rm ff}$, for the gas that forms the second sink particle. Periods when 
the gas is cooling are indicated in blue, while periods when the gas is heating are
indicated in red. 
Lower line: ratio of $t_{\rm thermal}$ to the orbital timescale, $t_{\rm orbital}$, for the
same set of SPH particles
(c) Temperature evolution of the gas that forms the second sink
(d) Density evolution of the gas that forms the second sink
\label{fig:thermo}}
\end{figure}

There are two main criteria that an accretion disk must satisfy before it is able to
undergo gravitational fragmentation -- the Toomre criterion discussed in the main article, 
and the Gammie criterion, which states that the thermal timescale of the gas in the
disk must be a small fraction of the orbital timescale \cite{gammie01}
\begin{equation}
t_{\rm thermal} \leq 3 \Omega^{-1} \equiv \frac{3}{2\pi} t_{\rm orbital},
\end{equation}
where $t_{\rm thermal} = e / \Lambda$, where $e$ is the internal energy density and
$\Lambda$ is the volumetric cooling (or heating) rate, $\Omega$ is the rotation frequency 
of the disk, and  $t_{\rm orbital} = 2\pi / \Omega$. If the thermal timescale does not satisfy 
the Gammie criterion, then gravitationally collapsing gas in the disk will undergo a thermal bounce 
and then be sheared apart by the disk rotation, rather than continuing to collapse. It is
therefore important to understand the thermal evolution of the material in the
disk, and to verify that the gas which fragments does indeed satisfy the Gammie
criterion. 

To do this, we have harnessed the Lagrangian nature of SPH. We first identify the SPH 
particle that finds itself at the center of the region in the disk that collapses to form 
the second protostar. This is the particle that is turned into a sink particle, once it 
(and its neighbors) pass the tests described in Section~\ref{method}. We then go back to a point in 
the simulation just after the central protostar is formed, and follow the evolution of 
this particle by computing the mean properties of the gas as seen by the particle and its 
50 nearest neighbors at many different instances in the life of the disk. It is this evolution that 
is shown in Figure S\ref{fig:thermo}.

We found that the gas which formed our second sink particle underwent several orbits
in the disk prior to becoming gravitationally unstable, passing into and out of the
spiral density wave pattern present in the disk. Gas entering the spiral density wave 
pattern was compressed, while gas exiting it was rarefied, as can be seen from panels
3 and 4 of Figure S\ref{fig:thermo}. During periods of compression, the dynamical
heat input (red circles in panel (a) of Figure S\ref{fig:thermo}) was balanced primarily 
by radiative cooling through H$_{2}$  line emission (pale blue line in panel (a) of Figure 
S\ref{fig:thermo}), even though the strongest of the H$_{2}$ lines all had significant 
optical depths at this point in the calculation. In comparison, cooling from H$_{2}$
collision-induced emission (CIE; blue dashed line in panel (a) of Figure S\ref{fig:thermo})
made only a minor contribution to the dissipation of energy from the disk throughout most
of the simulation.  The other cooling process included in the Figure, H$_{2}$ collisional 
dissociation cooling (dark blue line in panel (a) of Figure S\ref{fig:thermo}), is generally
negligible for the material in the disk. However, the exponential temperature dependence of
the collisional dissociation rate means that it rapidly becomes more important as the 
temperature rises. 

Once runaway gravitational collapse sets in at $t \sim 90$~yr, the relative importance
of the three main cooling processes changes. As the density and temperature of the gas 
increase, all three processes yield higher volumetric cooling rates. However, the rate 
of increase in the H$_{2}$ line cooling rate is slow, owing to the effects of the high
line opacity, and so it is quickly overtaken by CIE cooling. At the time of the last output
dump before the formation of the sink particle, H$_{2}$ collisional dissociation cooling
has less influence than H$_{2}$ line cooling or CIE cooling, and the gas remains almost
entirely molecular, with an atomic hydrogen fraction of order $10^{-3}$. However, it is
plain that the importance of H$_{2}$ collisional dissociation cooling is increasing rapidly
at this point, and we expect it to become the dominant process limiting the rate of
increase of the temperature of the collapsing gas at higher densities. This is important, as 
this is a non-radiative process, and hence one that will be unaffected by the increasing 
opacity of the collapsing gas. Although we do not follow the details of the collapse beyond 
our sink particle creation density, it is reasonable to assume that the collapse will
proceed in a quasi-isothermal fashion for as long as there remains H$_{2}$ available
to dissociate, just as in simulations of present-day star formation \cite{larson69,mi00},
or high dynamical range simulations of primordial star formation \cite{yoh08}.
Only once the H$_{2}$ content of the collapsing gas is exhausted will the collapse become
adiabatic. 

We can also draw an important conclusion from Figure S\ref{fig:thermo} regarding the
importance of accretion luminosity heating in determining the temperature evolution 
of the gravitationally unstable gas. It is clear from panel (a) of Figure S\ref{fig:thermo}
that the accretion luminosity generally plays only a minor role in  heating the gas,
in comparison to compressional and viscous heating in the disk, dominating only when
these terms are small or absent. Since our numerical treatment of the accretion luminosity 
heating is designed to maximize its effects, we can be confident of finding the same result 
were we to use a more accurate treatment of the accretion luminosity feedback. 
We can therefore also be confident that radiative feedback from
the first star to form does not suppress fragmentation in the disk at this stage in
the lifetime of the system. We cannot, of course, rule out significant feedback effects
later in the lifetime of the disk, when the mass of the central sink will be larger, 
but these lie beyond the scope of our present study.

As far as the Gammie criterion is concerned, we can see from panel (b) of Figure
S\ref{fig:thermo} that this is satisfied for almost the whole period plotted. 
Typically, $t_{\rm thermal} \sim 0.1 t_{\rm orbital}$, increasing above this value 
only occasionally. Most importantly, $t_{\rm thermal} / t_{\rm orbital}$ becomes
small once the gas begins to undergo runaway gravitational collapse, decreasing
to roughly $t_{\rm thermal} / t_{\rm orbital} \sim 0.01$ by the end of the simulation.
In contrast, we note that the gas in the disk finds the standard Rees \& Ostriker
criterion for ongoing gravitational collapse \cite{ro77}, namely that $t_{\rm ff} > t_{\rm thermal}$, 
more difficult to satisfy. It is this condition that helps to maintain the overall global 
stability  in the disk.

Finally, the results presented here allow us to understand why our conclusions
regarding the stability of Population III accretion disks differ significantly
from those of the previous analytical studies \cite{tm04,tb04,md05b}.
Figure S\ref{fig:thermo} demonstrates that H$_{2}$ line cooling plays a hugely
important role in the thermal balance of the disk, allowing the disk material
to remain relatively cold, with a temperature of $T \sim 1000$--2000~K. However,
this process was not included in any of these previous analytical studies.
They therefore find much higher equilibrium temperatures for the gas in the disk.
Neglect of H$_{2}$ bound-free opacity means that these studies predict
inner disk temperatures $T \sim 6000 \: {\rm K}$ or more, the temperature at
which H$^{-}$ ions first become a major source of opacity. At a temperature of
6000~K, the molecular content of the gas is negligible, and so the predicted
mean molecular weight of the gas in these models also differs by almost a 
factor of two from the value in our cold disks. Together, these effects lead
to a significantly higher predicted sound-speed for the disk, and hence also a
higher Toomre parameter $Q$. Our simulated disks are already marginally 
stable, and it is likely that a global increase in $Q$ by a factor of a few would
render them completely stable against fragmentation. The difference between 
the  results of these earlier analytical studies and our simulations can therefore
be understood as a direct consequence of the difference in the input physics, 
and underscores the importance of properly accounting for the H$_{2}$ line
cooling. 

\section{Potential impact of magnetic fields}

Magnetic fields in the early universe are usually assumed to be extremely weak and their potential influence on the gas dynamics is neglected in most numerical simulations of Population III star formation \cite{bromm09}. Upper limits on the field strength can be derived from the cosmic microwave background \cite{barr97}, from big-bang nucleosynthesis \cite{green69,mo70}, from modeling the reionization history \cite{sch08}, or from the $21\,$cm line \cite{sch09}. Typical values lie around $10^{-9}\,$G in comoving units. 

Although a variety of physical processes have been proposed for the creation of a seed magnetic field during inflation, for example via electroweak or QCD phase transitions \cite{gr01}, most studies conclude that the dominant contribution to the magnetic field strength comes from astrophysical processes after recombination. The Biermann battery has been proposed to generate fields of the order of $10^{-18}\,$G in the intergalactic medium at a redshift of $z \approx 20$ \cite{xu08}. This field could be amplified further via the Weibel instability in shocks \cite{ss03,lazar09}. The most likely mechanism for field amplification, however, is the dynamo process \cite{ps89}. Indeed, two recent studies \cite{sch10,sur10} conclude that the presence of a small-scale turbulent dynamo is able to increase the field strength by many orders of magnitude during the collapse of primordial gas clouds. Once a disk has formed, the large-scale dynamo and the magnetorotational instability may also become important \cite{tb04,sl06}. In addition, the presence of coherent fields could launch jets and outflows \cite{mach06,mach08}.

From studies of low-mass star formation in the present day, we know that the presence of magnetic fields with field strengths close to the equipartition value can effectively redistribute angular momentum via a process called magnetic braking \cite{mp79,hc09} and can thereby influence the fragmentation behavior \cite{ht08,hf08}. However, this requires the presence of coherent rotational motions in the disk. Calculations of massive star formation in the solar neighborhood with and without radiative feedback \cite{krum09,p10a,p10b} show that the disk around the central protostar quickly fragments to build up a cluster of stars, resulting in extremely complex and chaotic gas flows. It is likely that this will reduce the effectiveness with which magnetic fields can redistribute angular momentum and drive magnetic tower flows. As a consequence, the fragmentation behavior of the accretion disk around high-mass stars is not changed much by the presence of the field. As the disks studied in present-day high-mass star formation calculations are similar in nature to the Population III accretion disks modeled in our simulations, we feel confident that correctly accounting for the presence of primordial magnetic fields would not change our conclusions.

\end{document}